\documentclass[a4paper,11pt]{article}

\usepackage{jheppub, physics,hyperref,amsmath, float,array,adjustbox}
\usepackage[boxsize=6.5pt]{ytableau}
\usepackage{multirow}
 
\usepackage[T1]{fontenc}

\title{
\begin{center}
Chiral dualities for SQCD$_{3}$  with  D-type superpotential
\end{center}
}

\author[a]{Antonio Amariti}
\author[a,b]{and Davide Morgante}

\affiliation[a]{INFN, Sezione di Milano, Via Celoria 16, I-20133 Milano, Italy}
\affiliation[b]{Dipartimento di Fisica, Università degli studi di Milano, Via Celoria 16, I-20133, Milano, Italy}

\emailAdd{antonio.amariti@mi.infn.it}
\emailAdd{davide.morgante@mi.infn.it}

\abstract{We study dualities for $3d$ $\text{U}(N_{c})_k$  \textit{chiral} SQCD with $D_{n+2}$-type superpotential, with $n$ odd. 
We give a complete classification of such dualities in terms of  the number of fundamentals and anti-fundamentals and the Chern-Simons level. The classification is obtained by real mass and Higgs flows from non-chiral dualities and we check the consistency of the new non-chiral  dualities at the level of the partition function. We we also check that the complex phases appearing in the integral identities between the partition functions are consistent with the contact terms computed as quantum corrections to the effective Chern-Simons level. The $\text{SU}(N_{c})_k$ cases are recovered by gauging the topological symmetry from the $\text{U}(N_{c})$ dualities. Finally, we consider the case of $\text{USp}(2N_{c})_{2k}$  with two antisymmetric tensors and $D_{n+2}$-type superpotential.
}
\begin{document} 
\maketitle
\flushbottom
\section{Introduction}
\label{sec:intro}

The classification program of $3d$ $\mathcal{N}=2$ dualities is a fruitful field of research that boosted once localization techniques made powerful tools available \cite{Pestun:2007rz}.
Indeed after the discovery of mirror symmetry \cite{Intriligator:1996ex,deBoer:1996mp,deBoer:1996ck} and of Aharony duality \cite{Aharony:1997gp} (see also \cite{Karch:1997ux}), it took the community a decade to have another class of  examples of $3d$ $\mathcal{N}=2$ dualities  \cite{Giveon:2008zn,Niarchos:2008jb}. These examples were derived from the type IIB Hanany Witten (HW) setup \cite{Hanany:1996ie}, and they were motivated by the ABJ(M) results \cite{Aharony:2008ug,Aharony:2008gk}.

The matrix integral for the $3d$  $\mathcal{N}=2$  three sphere partition function, derived in full generality  \cite{Kapustin:2009kz,Jafferis:2010un,Hama:2010av,Hama:2011ea},
allowed to check the validity of these dualities and to define new ones, thanks to the possibility to engineer real mass and Higgs flows.
Such flows are ubiquitous in the analysis of $3d$ SUSY gauge theories, and they usually give rise to a chiral like field content, for the case of SQCD with unitary gauge group, i.e. there is a different number of fundamentals $N_f$ and on anti-fundamentals $N_a$. This difference requires a non zero Chern-Simons (CS) coupling for the invariance under large gauge transformations.
Surprisingly the  integral identities relating the three sphere partition functions of the new dualities \cite{Benini:2011mf} were already known to the mathematical community for $\text{U}(N_c)$ SQCD.
The classification of chiral dualities for $3d$ $\mathcal{N}=2$ SQCD was then extended to the $\text{SU}(N_c)$ case in \cite{Aharony:2014uya}.

A further goal has been to formulate new $3d$ $\mathcal{N}=2$ dualities analogous to the  the various generalization of $4d$ $\mathcal{N}=1$ Seiberg-like dualities.
The simplest extension, due to   \cite{Kutasov:1995ss}, regarded the case  of adjoint SQCD with $A_n$ type superpotential.
Two dualities have been proposed, with  \cite{Niarchos:2008jb}  or without  \cite{Kim:2013cma}  CS action and with $N_f=N_a$ \footnote{Actually more dualities have been obtained by adding monopole superpotentials  \cite{Amariti:2018wht}, but here we will not discuss such possibility.}.
The chiral case was then partially studied in \cite{Hwang:2015wna,Nii:2019qdx} for the $\text{U}(N_c)$ case, while a uniform treatment was then provided in \cite{Amariti:2020xqm} for both the $\text{U}(N_c)$ and $\text{SU}(N_c)$ case, generalizing the case without  adjoints (corresponding to the case of $A_1$ type superpotential.)

It is natural to wonder if these constructions can be generalized to the case of  SQCD with two adjoints interacting through   a $D_{n+2}$-type superpotential. The $4d$ duality was found in  \cite{Brodie:1996vx}, and the $3d$ analogous constructions have been discussed in \cite{Niarchos:2008jb} for the case with non-vanishing CS and more recently in \cite{Hwang:2018uyj} for the case with vanishing CS level.
This last case, obtained by dimensional reduction of the $4d$  duality of  \cite{Brodie:1996vx}, through the reduction scheme of   \cite{Aharony:2013dha}, reveals a novelty in the structure of the Coulomb branch of the $3d$ model, because of the presence of charge-two monopole operators in the superpotential of the magnetic phase.
Furthermore only the case of odd $n$ has been treated (see also \cite{Amariti:2022iaz} for further comments on the relation between 
the even $n$ case and real mass flows from the $\text{USp}(2N_c)$ SQCD with two antisymmetric version of this duality).

Here we start by considering  the $3d$ $\mathcal{N}=2$ two adjoint SQCD $\text{U}(N_c)_0$ duality of \cite{Hwang:2018uyj} and through a series of real mass and Higgs flows 
we generalize the web found in \cite{Benini:2011mf} and in \cite{Amariti:2020xqm} for the cases of SQCD and $A_n$ adjoint SQCD respectively.
Furthermore we gauge the topological symmetry  generalizing the construction to the $\text{SU}(N_c)$ case as well.
We corroborate the various steps of our derivation by reproducing the flow on the three sphere partition function, showing the cancellation of the divergent contributions, matching of the CS contact terms and proposing the new integral identities for the chiral dualities.
We conclude our analysis by studying the case of two antisymmetric $\text{USp}(2N_c)_{2k}$ SQCD with $D_{n+2}$-type superpotential, previously uncovered in the literature.

\section{Review}
\subsection{The $3d$ partition function} \label{sec:partFun}
In this section we give some basic results on the three-sphere partition function useful for our analysis. The partition function of a $3d$ $\mathcal{N}=2$ SQFT, computed through localization techniques on the squashed three-sphere $S^{3}_{b}$  \cite{Hama:2011ea},  corresponds to a matrix integral over a variable associated to  the real scalar of the vector multiplet in the Cartan of the gauge group.\\
The general structure of the partition function consists of classical contributions from the Fayet-Iliopulos (FI) and the CS terms in the action, and contributions coming from the one-loop determinants for the chiral and vector multiplets. If we consider a $3d$ $\mathcal{N}=2$ supersymmetric gauge theory with gauge group $G$ at CS level $k$, the $S^{3}_{b}$ partition function takes the following form
\begin{equation}
\begin{split}
	\mathcal{Z}_{G_{k}}(\mu_{a};\lambda) &= \frac{1}{\abs{W}} \int \prod_{i=1}^{\rank G}\frac{\dd{\sigma_{i}}}{\sqrt{-\omega_{1}\omega_{2}}}\exp(-i\pi k\sigma_{i}^{2}-i\pi\lambda\sigma_{i})\\
	&\times\prod_{I}\Gamma_{h}\qty(\omega\Delta_{I}+\rho_{I}(\sigma)+\tilde{\rho}_{I}(\mu))\;\cdot\prod_{\alpha\in G_{+}}\Gamma_{h}^{-1}(\pm\alpha(\sigma))
\end{split}
\label{eq:Zgeneral}
\end{equation}
where $\mu_{a}$ are real parameters associated to the flavour symmetry, while $\tilde{\rho}(\mu)$ and $\rho(\sigma)$ are the weights of the flavour and gauge symmetry respectively. The $\alpha$ are the positive roots of the gauge symmetry and they parametrize the contributions from the one-loop determinant of the vector multiplet.
The contribution of the FI, corresponding to the real mass for the topological symmetry $U(1)_J$, is parameterized by 
$\lambda$.
The R-charges of the chiral fields are parameterized by  $\Delta_{I}$. The gaussian factor corresponds to the CS level $k$. The normalization $\abs{W}$ is the order of the Weyl group of $G$.

In our notation, the one-loop dererminants are given in terms of hyperbolic Gamma functions which can be written as the following infinite product
\begin{equation}
	\Gamma_{h}(x) = e^{\frac{i\pi}{2\omega_{1}\omega_{2}}\qty((x-\omega)^{2}-\frac{\omega_{1}^{2}+\omega_{2}^{2}}{12})}\prod_{j=0}^{\infty}\frac{1-e^{\frac{2\pi i}{\omega_{1}}(\omega_{2}-x)}e^{\frac{2\pi i \omega_{2}j}{\omega_{1}}}}{1-e^{-\frac{2\pi i}{\omega_{2}}}e^{-\frac{2\pi i \omega_{2}j}{\omega_{2}}}}
\end{equation}
where $\omega_{1}=ib$, $\omega_{2}=i/b$ and $b$ is the squashing parameter of the three-sphere $S^{3}_{b}$ which is defined by $b^{2}(x_{1}^{2}+x_{2}^{2})+b^{-2}(x_{3}^{2}+x_{4}^{2})=1$; the $\omega$ parameter is defined as $2\omega=\omega_{1}+\omega_{2}$. We will often use the compound notation where 
\begin{equation}
	\Gamma_{h}(x;y) \equiv \Gamma_{h}(x)\Gamma_{h}(y),\qquad\Gamma_{h}(\pm x) = \Gamma_{h}(x)\Gamma_{h}(-x).	
\end{equation}

The hyperbolic Gamma function obeys useful identities that are going to play an essential role in our analysis. The first is the inversion formula 
\begin{equation}
	\Gamma_{h}(2\omega-x)\Gamma_{h}(x)=1
	\label{eq:inversion}
\end{equation}
which in field theory corresponds to integrating out fields appearing in the superpotential through holomorphic mass terms. The second one gives  the asymptotic behavior of the hyperbolic Gamma function
\begin{equation}
	\lim_{\abs{x}\rightarrow\infty}\Gamma_{h}(x)=e^{-\frac{i\pi}{2}\operatorname{sign}(x)(x-\omega)^{2}}
	\label{eq:asymptGamma}
\end{equation}
and it corresponds in field theory to integrating out a massive field with a large real mass term.

Focusing on the chiral models of our interest, the partition function of a $\text{U}(N_{c})_{k}$ theory with $N_{f}$ fundamentals, $N_{a}$ anti-fundamentals and two adjoints $X,Y$, at CS level $k$ is
\begin{equation}
\begin{split}
	\mathcal{Z}_{\text{U}(N_{c})_{k}}^{(N_{f},N_{a})}(\vec\mu;\vec\nu;\tau_{X};\tau_{Y};\lambda)&=\frac{\Gamma_{h}(\tau_{X})^{N_{c}}\Gamma_{h}(\tau_{Y})^{N_{c}}}{N_{c}!\sqrt{-\omega_{1}\omega_{2}}^{N_{c}}}\int\prod_{i=1}^{N_{c}}\dd{\sigma_{i}}\exp\qty(-i\pi\lambda\sigma_{i} -i\pi k\sigma_{i}^{2})\\
	&\times\prod_{1\le i < j \le N_{c}} \prod_{\beta=X,Y}\frac{\Gamma_{h}\big(\tau_{\beta}\pm(\sigma_{i}-\sigma_{j})\big)}{\Gamma_{h}\big(\pm(\sigma_{i}-\sigma_{j})\big)}\\
	&\times\prod_{i=1}^{N_{c}} \bigg( \prod_{a=1}^{N_{f}}\Gamma_{h}(\mu_{a}+\sigma_{i})\cdot\prod_{b=1}^{N_{a}}\Gamma_{h}(\nu_{b}-\sigma_{i}) \bigg).
\end{split}
\end{equation}
The parameters $\mu_{a},\nu_{a}$ refer to the real masses of the fundamentals and anti-fundamentals while the $\tau_{X,Y}$ are the real masses of the adjoints. 

The chiral $\text{SU}(N_{c})_{k}$ case can be recovered by the $\text{U}(N_{c})_{k}$ one by gauging the topological $\text{U}(1)_{J}$ symmetry which at the level of the partition function amounts to adding a factor $\frac{1}{2}e^{i\pi\lambda N_{c}m_{B}}$ and integrating over $\lambda$, imposing the tracelessness condition on the adjoint fields \cite{Amariti:2020xqm,Benvenuti:2020gvy}
\begin{equation}
\begin{split}
	\mathcal{Z}_{\text{SU}(N_{c})_{k}}^{(N_{f},N_{a})}(\vec\mu;\vec\nu;\tau_{X};\tau_{Y})&=\frac{\Gamma_{h}(\tau_{X})^{N_{c}-1}\Gamma_{h}(\tau_{Y})^{N_{c}-1}}{N_{c}!\sqrt{-\omega_{1}\omega_{2}}^{N_{c}}}\int\prod_{i=1}^{N_{c}}\dd{\sigma_{i}} \delta\qty(\sum_{i=1}^{N_{c}}\sigma_{i})\exp\qty(-i\pi k\sigma_{i}^{2})\\
	&\times\prod_{1\le i < j \le N_{c}}\prod_{\beta=X,Y}\frac{\Gamma_{h}\big(\tau_{\beta}\pm(\sigma_{i}-\sigma_{j})\big)}{\Gamma_{h}\big(\pm(\sigma_{i}-\sigma_{j})\big)}\\
	&\times\prod_{i=1}^{N_{c}} \bigg(\prod_{a=1}^{N_{f}}\Gamma_{h}(\mu_{a}+m_{B}+\sigma_{i})\cdot\prod_{b=1}^{N_{a}}\Gamma_{h}(\nu_{b}-m_{B}-\sigma_{i})\bigg).
\end{split}
\end{equation}

\subsection{Non-chiral $3d$ dualities with adjoint matter}

We start our analysis from the $3d$ $\mathcal{N}=2$ duality for $\text{U}(N_c)_0$ SQCD with $N_f$ pairs of fundamentals and anti--fundamentals and two adjoints interacting through a $D_{n+2}$-type superpotential.
The duality has been obtained in \cite{Hwang:2018uyj}, from the circle reduction of the $4d$ duality of \cite{Brodie:1996vx},
by following the prescription of \cite{Aharony:2013dha}.
In the $3d$ limit the duality is characterized by the unusual presence of superpotential interactions involving (dressed) monopole operators of charge two.
The duality relates

\begin{itemize}
	\item $3d$ $\mathcal{N}=2$ $\text{U}(N_{c})_0$ SQCD with $N_{f}$ flavours $Q,\tilde{Q}$ with two adjoints fields $X,Y$ and  superpotential
	\begin{equation}
		\mathcal{W}_{\text{ele}} = \Tr X^{n+1}+\Tr XY^{2}
		\label{eq:Uele}
	\end{equation}
	with $n$ odd.
	\item $3d$ $\mathcal{N}=2$ $\text{U}(\tilde{N}_{c})_{0}$ SQCD with $\tilde{N}_{c}=3nN_{f}-N_{c}$, $N_{f}$ dual flavours $q,\tilde{q}$ and two adjoint fields  $x,y$ interacting through the superpotential
	\begin{equation}
	\begin{split}
		\mathcal{W}_{\text{mag}}&=\Tr x^{n+1}+\Tr xy^{2}+\sum_{j=0}^{n-1}\sum_{\ell=0}^{2}\Tr\qty(\mathcal{M}^{j,\ell}qx^{n-1-j}y^{2-\ell}\tilde{q})\\
		&+\sum_{\substack{j=0,\ldots,n-1\\\ell=0,1,2\\j\ell=0}}V^{\pm}_{j,\ell}\widetilde{V}^{\pm}_{n-j,2-\ell}+\sum_{q=0}^{\frac{n-3}{2}}W^{\pm}_{q}\widetilde{W}^{\pm}_{\frac{n-3}{2}-q} \, ,
		\label{eq:Umag}
	\end{split}
	\end{equation}
\end{itemize}
where singlets $\mathcal{M}^{j,\ell}$ are dual to the dressed mesons of the electric theory $QX^{j}Y^{\ell}\tilde{Q}$, the $V^{\pm}_{j,\ell}$ and $W^{\pm}_{q}$ are the monopole operators of the electric theory with topological charges $\pm 1$ and $\pm 2$ respectively, acting as singlets in the magnetic phase.
Observe that such monopole operators, defined through radial quantization from states on $S^2$ that carry a non--trivial magnetic flux background, are mapped to the states $\Tr X^j Y^\ell |\pm1,0\dots,0 \rangle$ and
$\Tr X^{2q} |\pm2,0\dots,0 \rangle$ respectively.

The  global symmetry  is $\text{SU}(N_{f})_{L}\times \text{SU}(N_{f})_{R}\times \text{U}(1)_{A}\times\text{U}(1)_{J}\times \text{U}(1)_{R}$, where $\text{U}(1)_{A}$ is the axial symmetry, $\text{U}(1)_{J}$ is the topological symmetry  and 
$\text{U}(1)_{R}$ is the  R-symmetry.
The various fields transform as in table \ref{tab:Un}. 
\begin{table}
\centering
\renewcommand{\arraystretch}{1.2}
\caption{Matter content of electric (upper) and magnetic (lower) $\text{U}(N_{c})_{0}$ theories.}
\vspace{5pt}
\ytableausetup{centertableaux}
\adjustbox{max width=\columnwidth}{
\begin{tabular}{||c|c|c|c|c|c|c|c||}
	\cline{2-8}
	\multicolumn{1}{c|}{}&\multicolumn{2}{c|}{Gauge}&\multicolumn{5}{c||}{Global}\\
	\hline
	Field & U$(N_{c})$ & U$(\tilde N_{c})$ & SU$(N_{f})_{L}$ & SU$(N_{f})_{R}$ & U$(1)_{A}$ & U$(1)_{J}$ & U$(1)_{R}$\\
	\hline
	$Q$ & $\Box$ & $1$ & $\Box$ & $1$ & $1$ & $0$ & $r_{Q}$\\
	$\tilde Q$ & $\overline\Box$ & $1$ & $1$ & $\overline\Box$ & $1$ & $0$ & $r_{Q}$\\
	$X$ & Adj & $1$ & $1$ & $1$ & $0$ & $0$ & $\frac{2}{n+1}$\\
	$Y$ & Adj & $1$ & $1$ & $1$ & $0$ & $0$ & $\frac{n}{n+1}$\\
	$V_{j\ell}^{\pm}$ & 1 & $1$ & $1$ & $1$ & $-N_{f}$ & $\pm 1$ & $(1-r_{Q})N_{f}+\frac{2j+n\ell-(N_{c}-1)}{n+1}$\\
	$W_{q}^{\pm}$ & 1 & $1$ & $1$ & $1$ & $-2N_{f}$ & $\pm 2$ & $2(1-r_{Q})N_{f}+\frac{2+4q-2(N_{c}-1)}{n+1}$\\
	\hline\hline
	$q$ & $1$ & $\overline\Box$ & $1$ & $\overline\Box$ & $-1$ & $0$ & $\frac{2-n}{n+1}-r_{Q}$\\
	$\tilde q$ & $1$ & $\Box$ & $\Box$ & $1$ & $-1$ & $0$ & $\frac{2-n}{n+1}-r_{Q}$\\
	$x$ & $1$ & Adj & $1$ & $1$ & $0$ & $0$ & $\frac{2}{n+1}$\\
	$y$ & $1$ & Adj & $1$ & $1$ & $0$ & $0$ & $\frac{n}{n+1}$\\
	$\mathcal{M}_{j\ell}$ & $1$ & $1$ & $\Box$ & $\overline\Box$ & $2$ & $0$ & $2r_{Q}+\frac{2j+n\ell}{n+1}$\\
	$\tilde V_{j\ell}^{\pm}$ & 1 & $1$ & $1$ & $1$ & $N_{f}$ & $\mp 1$ & $(r_{Q}-1)N_{f}+\frac{2j+n\ell+(N_{c}+1)}{n+1}$\\
	$\tilde W_{q}^{\pm}$ & 1 & $1$ & $1$ & $1$ & $2N_{f}$ & $\mp 2$ & $2(r_{Q}-1)N_{f}+\frac{2+4q+2(N_{c}+1)}{n+1}$\\
	\hline
\end{tabular}}
\label{tab:Un}
\end{table}

The $4d/3d$ reduction of the  duality of \cite{Brodie:1996vx}  has been recently studied also in \cite{Amariti:2022iaz} by circle reduction of the conjectured identity between the $4d$ superconformal indices. The final result, once the divergent contributions  between the electric and the magnetic side of the identity have been matched and canceled, corresponds to the identity that reproduces the $3d$ $\mathcal{N}=2$ duality of \cite{Hwang:2018uyj} on the squashed three sphere partition function.
The identity  is

\begin{equation}
\begin{split}
	\mathcal{Z}_{\text{U}(N_{c})}^{N_{f}}(\mu_{a};\nu_{a};\tau_{X};\tau_{Y};\lambda)&=\mathcal{Z}_{\text{U}(\tilde{N}_{c})}^{N_{f}}(\tau_{X}-\tau_{Y}-\nu_{a};\tau_{X}-\tau_{Y}-\mu_{a};\tau_{X};\tau_{Y};-\lambda)\\
	&\times\prod_{j=0}^{n-1}\prod_{\ell=0}^{2}\prod_{a,b=1}^{N_{f}}\Gamma_{h}(j\tau_{X}+\ell\tau_{Y}+\mu_{a}+\nu_{b})\\
	&\times  \!\!\!\! \!\!\!\!\prod_{\substack{j=0,\ldots,n-1\\\ell=0,1,2\\j\ell=0}}
	\!\!\!\! \!\!\!\!\Gamma_{h}\qty(\pm \frac{\lambda}{2}+N_{f}\omega-\frac{N_{c}-1}{2}\tau_{X}-\frac{1}{2}\sum_{a=1}^{N_{f}}(\mu_{a}+\nu_{a})+j\tau_{X}+\ell\tau_{Y})\\
	&\times\prod_{q=0}^{\frac{n-3}{2}}\Gamma_{h}\qty(\pm\lambda+2N_{f}\omega+(N_{c}-1)\tau_{X}\!\!-\!\!\sum_{a=1}^{N_{f}}(\mu_{a}+\nu_{a})+(2q+1)\tau_{X}).
	\label{eq:dualU}
\end{split}
\end{equation}

The parameters associated to the $N_f$ fundamentals and anti-fundamentals satisfy the constraint $\sum_{a=1}^{N_{f}}\mu_{a}=\sum_{a=1}^{N_{f}}\nu_{a}=N_{f}m_{A}$.
The parameters associated to the adjoint are fixed as
\begin{equation}
	\tau_{X}=\frac{2\omega}{n+1},\qquad \tau_{Y}=\frac{n \omega}{n+1}
\end{equation}
reflecting the constraints imposed by the superpotential (\ref{eq:Uele}).

\subsection{Classification of chiral-dualities}

We conclude our review by surveying  the case of chiral dualities for ordinary $3d$ $\mathcal{N}=2$ SQCD
worked out in \cite{Benini:2011mf}.
These dualities are characterized by a different number of fundamentals $N_f$ and anti-fundamentals $N_a$, 
and by a CS level $k$. By comparing $\Delta F\equiv|N_f-N_a|$ and $2k$ three different cases have been identified.

For historical reasons the classification proposed in  \cite{Benini:2011mf} for such dualities reflects the one worked out in the mathematical literature
for the hyperbolic integral identities, corresponding to the matching of the three sphere partition functions (see for example \cite{vanDeBult}).
In this case the chiral SQCD models are labelled by two non-negative integers $\comm{\mathbf{p}}{\mathbf{q}}$. These integral identities are related to the physics of CS theories with chiral matter. 

The relation between the integers $\comm{\mathbf{p}}{\mathbf{q}}$ and the physical quantities 
can be made explicit  by defining these integers in terms of the effective CS level of a $\text{U}(N_{c})_{k}$ theory 
\begin{equation}
	k_{\pm}=k\pm\frac{1}{2}(N_{f}-N_{a}).
	\label{eq:effCS}
\end{equation}
According to the signs of $k_{\pm}$, there are four possible definitions
\begin{align}
\label{idkpkm}
	\comm{\mathbf{p}}{\mathbf{q}}_{a}&\equiv\comm{-k_{+}}{-k_{-}}_{a},	&\comm{\mathbf{p}}{\mathbf{q}}^{*}_{a}\equiv&\comm{-k_{+}}{k_{-}}_{a}^{*}, \nonumber \\
	\comm{\mathbf{p}}{\mathbf{q}}_{b}&\equiv\comm{k_{+}}{k_{-}}_{b}, &\comm{\mathbf{p}}{\mathbf{q}}^{*}_{b}\equiv&\comm{k_{+}}{-k_{-}}^{*}_{b}
\end{align}
where the theory type $a,b$ is chosen such that $\mathbf{p},\mathbf{q}>0$.

This means that for any choice of $k$, $N_{f}$ and $N_{a}$ one has to compute $k_{\pm}$  using (\ref{eq:effCS}) and then
one has to select in (\ref{idkpkm}) the one with both  $\mathbf{p}$ and $\mathbf{q}$ positive.
The flip of the sign of the CS term under duality imposes also that the dual of an $a$-theory is a $b$-theory and viceversa \cite{Benini:2011mf}. 

We survey the classification the dualities of \cite{Benini:2011mf} following this notation and based on the difference between $\Delta F$ and $2k$ (with $k>0$, the case of $k<0$ can be studied analogously). In each case the electric theories are $\text{U}(N_{c})_{k}$ SQCD with $N_{f}$ fundamentals and $N_{a}$ anti-fundamentals and vanishing superpotential. Depending on the value of $\comm{\mathbf{p}}{\mathbf{q}}$ one has
\begin{equation}
\begin{split}
	&\comm{\mathbf{p}}{\mathbf{0}}\qquad \Delta F = 2k \qquad N_{a}<N_{f},\\
	&\comm{\mathbf{p}}{\mathbf{q}}\qquad \Delta F<2k \qquad N_{a}\neq N_{f},\\
	&\comm{\mathbf{p}}{\mathbf{q}}^{*}\qquad \!\!\Delta F>2k \qquad N_{a}\neq N_{f}.
\end{split}
\end{equation}
The gauge group of the dual magnetic chiral SQCD is 
\begin{equation}
\begin{split}
	&\comm{\mathbf{p}}{\mathbf{0}}\qquad\text{U}(N_{f}-N_{c})_{-k}\;, \\
	&\comm{\mathbf{p}}{\mathbf{q}}\qquad\text{U}\Big(\frac{1}{2}(N_{f}+N_{a})+\abs{k}-N_{c}\Big)_{-k}\;,\\
	&\comm{\mathbf{p}}{\mathbf{q}}^{*}\qquad\!\!\text{U}\big(\max(N_{a},N_{f})-N_{c}\big)_{-k}
\end{split}
\end{equation}
with $N_{a}$ fundamentals and $N_{f}$ anti-fundamentals. In the last two cases the dual superpotential is given by
\begin{equation}
	\mathcal{W}_{\text{mag}}=Mq\tilde{q}
\end{equation}
while in the $\comm{\mathbf{p}}{\mathbf{0}}$ case only there is an additional singlet $T_{+}$ in the magnetic phase and the superpotential is given by
\begin{equation}
	\mathcal{W}_{\text{mag}}=Mq\tilde{q} +T_{+}t_{-}
\end{equation}
where $T_{+}$ is dual to the electric monopole.

An analogous description holds for the $\text{SU}(N_{c})$ cases (see \cite{Aharony:2014uya,Amariti:2020xqm}).
Furthermore the discussion has been extended to the case of adjoint SQCD with $A_n$-type superpotential
and unitary gauge group \cite{Hwang:2015wna,Amariti:2020xqm}.
In the following we will discuss the generalization to the case of two adjoint SQCD with $D_{n+2}$-type superpotential
and unitary gauge group.
\section{Dualities for $\text{U}(N_{c})$ chiral SQCD with two adjoints}
In this section, we study the chiral limit of the $\text{U}(N_{c})_{0}$ duality studied in \cite{Hwang:2018uyj}. We will use the above-mentioned notation to differentiate the various theories with the addition of the subscript $\comm{\mathbf{p}}{\mathbf{q}}_{X,Y}$ to underline the presence of two adjoints, similar to the notation of \cite{Amariti:2020xqm}. 

We introduce real mass flows on the electric side by turning on background fields for the flavour symmetry and giving large vacuum expectation values to the scalars in the vector multiplet of the flavour symmetry. This flow will lead in the IR to $\comm{\mathbf{p}}{\mathbf{q}}_{X,Y}$ theories. Then we turn on background fields for the gauge symmetry on the magnetic side and consider large vacuum expectation values to the scalars.\\
This procedure is  rephrased on the partition function (\ref{eq:dualU}) by considering consistent assignments of shifts on the parameters associated to the flavour and to the gauge symmetry. 
For large shifts the asymptotic of the integral identities gives new finite identities for the partition functions after factoring  and canceling out 
the divergent contributions between the electric and magnetic phases.
\subsection{The $\comm{\mathbf{p}}{\mathbf{p}}_{X,Y}$ case}
The  $\comm{\mathbf{p}}{\mathbf{p}}_{X,Y}$ duality (corresponding to the one studied in \cite{Niarchos:2008jb}) is obtained 
from the $\comm{\mathbf{0}}{\mathbf{0}}_{X,Y}$ duality with $N_{f}+k$ flavours by assigning a positive large real mass to $k$ of them.
In the magnetic phase $k$ dual quarks and anti-quarks are shifted with opposite signs while $2n(k^2+2 N_f k)$ components of the (dressed) mesons acquire a large mass  accordingly.
 In the IR, this will lead to the following duality:
\begin{itemize}
	\item $\text{U}(N_{c})_{k}$ SQCD with $N_{f}$ fundamentals and anti-fundamentals $Q,\tilde{Q}$ and two adjoints $X,Y$ interacting through the superpotential
	\begin{equation}
		\mathcal{W}_{\text{ele}} = \Tr X^{n+1}+\Tr XY^{2}.
	\end{equation}
	\item $\text{U}(\tilde{N}_{c})_{-k}$ SQCD, with $\tilde{N}_{c}=3n(N_{f}+\abs{k})-N_{c}$,  $N_{f}$ fundamentals and anti-fundamentals $q,\tilde{q}$ and two adjoints $x,y$, interacting through the superpotential
	\begin{equation}
		\mathcal{W}_{\text{mag}}=\Tr x^{n+1}+\Tr xy^{2}+\sum_{j=0}^{n-1}\sum_{\ell=0}^{2}\Tr\qty(\mathcal{M}^{j,\ell}qx^{n-1-\ell}y^{2-l}\tilde{q}).
	\end{equation}
\end{itemize}
The two theories acquire a CS level $k$ and $-k$ respectively.
The CS term lifts the Coulomb branch of the $\text{U}(N_{c})$ model. It reflects in the dual side to integrate out
the singlets corresponding to the  monopole of the electric phase.

To reproduce the duality on the partition function, we start from the equality (\ref{eq:dualU}) and consider the following shifts of the real masses
\begin{equation}
\begin{cases}
	m_{A}\rightarrow m_{A}+\frac{k}{N_{f}+k}s\\
	m_{a}\rightarrow m_{a}-\frac{k}{N_{f}+k}s& a=1,\ldots,N_{f}\\
	m_{a}\rightarrow m_{a}+\frac{N_{f}}{N_{f}+k}s& a=1,\ldots,k\\
	n_{a}\rightarrow n_{a}-\frac{k}{N_{f}+k}s& a=1,\ldots,N_{f}\\
	n_{a}\rightarrow n_{a}+\frac{N_{f}}{N_{f}+k}s& a=1,\ldots,k
\end{cases}
\label{eq:shiftUpp}
\end{equation}
where we split the abelian axial part, $m_{A}$, of the real masses for the flavour symmetry from its non-abelian part $m_{a},n_{a}$.

At this level, when the shift is finite, the equality (\ref{eq:dualU}) still holds. To reproduce the flow, we need study the large $s$ limit on the partition functions by making use of the asymptotic behavior of the hyperbolic Gamma function (\ref{eq:asymptGamma}). One needs to be careful when taking this limit since an infinite shift in the variables makes the integrals divergent. Therefore we need check that in the limit the leading saddle point contributions cancel between the electric and magnetic partition functions \cite{Amariti:2014iza}. We are left then with the equality between
\begin{equation}
\begin{split}
	\mathcal{Z}_{\text{ele}}=&\mathcal{Z}_{\text{U}(N_{c})_{k}}^{(N_{f},N_{f})}(\mu_{a};\nu_{a};\tau_{X};\tau_{Y};\lambda)
\end{split}
\end{equation}
and 
\begin{equation}
\begin{split}
	\mathcal{Z}_{\text{mag}} &= e^{i\pi\phi}e^{-\frac{3i\pi}{4}n\lambda^{2}}\mathcal{Z}_{\text{U}(\tilde{N}_{c})_{-k}}^{(N_{f},N_{f})}(\tau_{X}-\tau_{Y}-\nu_{a};\tau_{X}-\tau_{Y}-\mu_{a};\tau_{X};\tau_{Y};-\lambda)\\
	&\times\prod_{j=0}^{n-1}\prod_{	\ell=0}^{2}\prod_{a,b=1}^{N_{f}}\Gamma_{h}(j\tau_{X}+\ell\tau_{Y}+\mu_{a}+\nu_{b}),
\end{split}
\end{equation}
where $\mu_{a}=m_{a}+m_{A}$ and $\nu_{b}=n_{b}+m_{A}$, satisfying the constraint $\sum_{a=1}^{N_{f}}\mu_{a}=\sum_{b=1}^{N_{f}}\nu_{b}=N_{f} m_{A}$.\\

There is a non-trivial complex exponential phase
in the identity between $\mathcal{Z}_{\text{ele}}$ and $\mathcal{Z}_{\text{mag}} $. This phase
is  essential for matching the partition functions of the two dual theories.  It was shown  \cite{Closset:2012vp,Closset:2012vg} that the exponents are related to CS contact terms in two-point functions of global symmetry currents.
The complex phase $\phi$ in this case has the following form
\begin{equation}
\begin{split}
	\phi&=2N_{f}m_{A}\tau_{Y}(\tilde{N}_{c}-2N_{c}+3N_{f}+3k(n-1))-\frac{\tau_{X}^{2}}{8}\\
	&-\frac{1}{4}\tau_{X}\tau_{Y}\Big((1+n+n^{2})+6N_{f}^{2}+k^{2}+6N_{c}^{2}-4N_{f}(k+3N_{c})+4(N_{c}+\tilde{N}_{c})(N_{f}+k)\\
	&-(12k^{2}+12N_{f}N_{c}+18kN_{c})n+6(N_{f}^{2}+4N_{f}k+4k^{2})n^{2}\Big)+3n N_{f}m_{A}^{2}(k-N_{f})\\
	&+\frac{3}{2}kn\sum_{a=1}^{N_{f}}(m_{a}^{2}+n_{a}^{2}).
\end{split}
\end{equation}
This phase can be reproduced from the computation of the contact terms by a linear combination of $\Delta k_{ij}$, where the indices run over the abelian symmetries. 
Observe that in this case we have only a discrepancy in $\Delta k_{rr}$ with respect to the result red from the exponent in the partition function. This is nevertheless unphysical because it only acts as a pure phase in the identity between 
$\mathcal{Z}_{\text{ele}} $ and $\mathcal{Z}_{\text{mag}} $.

\subsection{The $\comm{\mathbf{p}}{\mathbf{q}}_{X,Y}$ case}
The flow to the $\comm{\mathbf{p}}{\mathbf{q}}_{X,Y}$ duality is obtained  starting from the $\comm{\mathbf{0}}{\mathbf{0}}_{X,Y}$ $\text{U}(N_{c})_{0}$ duality with $N_{f}$ flavours by assigning a positive large real mass to $N_{f}-N_{f_{1}}$ fundamentals and $N_{f}-N_{f_{2}}$ anti-fundamentals. In the IR, this will lead the following duality:
\begin{itemize}
	\item $\text{U}(N_{c})_{k}$ SQCD with $N_{f_{1}}$ fundamentals and $N_{f_{2}}$ anti-fundamentals $Q,\tilde{Q}$, two adjoints $X$ and $Y$ interacting through the superpotential
	\begin{equation}
		\mathcal{W}_{\text{ele}} = \Tr X^{n+1}+\Tr XY^{2}.
	\end{equation}
	\item $\text{U}(\tilde{N}_{c})_{-k}$ SQCD, with $\tilde{N}_{c}=3nN_{f}-N_{c}$,  $N_{f_{2}}$ fundamentals $q$,  $N_{f_{1}}$ anti-fundamentals $\tilde{q}$ and two adjoint fields $x,y$, interacting through the superpotential
	\begin{equation}
		\mathcal{W}_{\text{mag}}=\Tr x^{n+1}+\Tr xy^{2}+\sum_{j=0}^{n-1}\sum_{\ell=0}^{2}\Tr\qty(\mathcal{M}^{j,\ell}qx^{n-1-j}y^{2-\ell}\tilde{q}).
	\end{equation}
\end{itemize}
The CS levels of the two phases are given by $k=N_{f}-\frac{1}{2}(N_{f_{1}}+N_{f_{2}})$ and $-k$ respectively. The Coulomb branch of the electric phase is lifted and in the dual phase the dressed electric monopoles acting as singlets are now massive and we integrated them out.

To reproduce the duality on the partition function, we start from the equality (\ref{eq:dualU}) and consider the following shifts in the real masses
\begin{equation}
\begin{cases}
	m_{A}\rightarrow m_{A}+\frac{2N_{f}-N_{f_{1}}-N_{f_{2}}}{2N_{f}}s\\[3pt]
	m_{a}\rightarrow m_{a}-\frac{N_{f}-N_{f_{1}}}{N_{f}}s& a=1,\ldots,N_{f_{1}}\\
	m_{a}\rightarrow m_{a}+\frac{N_{f_{1}}}{N_{f}}s& a=1,\ldots,N_{f}-N_{f_{1}}\\
	n_{a}\rightarrow n_{a}-\frac{N_{f}-N_{f_{2}}}{N_{f}}s& a=1,\ldots,N_{f_{2}}\\
	n_{a}\rightarrow n_{a}+\frac{N_{f_{2}}}{N_{f}}s& a=1,\ldots,N_{f}-N_{f_{2}}\\
	\sigma_{i}\rightarrow \sigma_{i}-\frac{N_{f_{1}}-N_{f_{2}}}{2N_{f}}s& i=1,\ldots,N_{c}\\
	\tilde\sigma_{i}\rightarrow \tilde\sigma_{i}-\frac{N_{f_{1}}-N_{f_{2}}}{2N_{f}}s& i=1,\ldots,3nN_{f}-N_{c}\\
	\lambda\rightarrow\lambda+(N_{f_{2}}-N_{f_{1}})s 
\end{cases}
\end{equation}
We study the limit of large $s$ as stated before, checking that the divergent contributions cancel between the electric and magnetic phases. We are left with the equality between
\begin{equation}
\begin{split}
	\mathcal{Z}_{\text{ele}}=&\mathcal{Z}_{\text{U}(N_{c})_{k}}^{(N_{f_{1}},N_{f_{2}})}\qty(\mu_{a};\nu_{b};\tau_{X};\tau_{Y};\hat\lambda)
\end{split}
\end{equation}
where
\begin{equation}
	\hat\lambda = \lambda+(N_{f_{1}}-N_{f_{2}})(m_{A}-\omega),
\end{equation}
and 
\begin{equation}
\begin{split}
	\mathcal{Z}_{\text{mag}} &= e^{i\pi\phi}e^{-\frac{3i\pi}{4}n\lambda^{2}}\mathcal{Z}_{\text{U}(\tilde{N}_{c})_{-k}}^{(N_{f_{2}},N_{f_{1}})}(\tau_{X}-\tau_{Y}-\nu_{a};\tau_{X}-\tau_{Y}-\mu_{a};\tau_{X};\tau_{Y};\tilde\lambda)\\
	&\times\prod_{j=0}^{n-1}\prod_{	\ell=0}^{2}\prod_{a=1}^{N_{f_{1}}}\prod_{b=1}^{N_{f_{2}}}\Gamma_{h}(j\tau_{X}+\ell\tau_{Y}+\mu_{a}+\nu_{b})\\
\end{split}
\end{equation}
where 
\begin{equation}
	\tilde\lambda = -\lambda-(N_{f_{1}}-N_{f_{2}})(m_{A}-\tau_{X}+\tau_{Y}+\omega).
\end{equation}
We set $\mu_{a}=m_{a}+m_{A}$ and $\nu_{b}=n_{b}+m_{A}$, satisfying  the constraint $\sum_{a=1}^{N_{f}}\mu_{a}=\sum_{b=1}^{N_{f}}\nu_{b}=N_{f} m_{A}$.\\
The complex exponent $\phi$, necessary for the equality between the partition functions, has the following form
\begin{equation}
\begin{split}
\label{phdav}
	\phi&=-\frac{3}{2}m_{A}\tau_{Y}\qty((1+n)(N_{f_{1}}+N_{f_{2}})(2k+N_{f_{1}}+N_{f_{2}})-2\tilde{N}_{c}(N_{f_{1}}+N_{f_{2}})+4N_{f_{1}}N_{f_{2}}(n-2))\\
	&-\frac{\tau_{X}^{2}}{8}-\frac{\tau_{X}\tau_{Y}}{4}\Big((1+n+n^{2})+11N_{f_{1}}N_{f_{2}}+6n(n-2)N_{f_{1}}N_{f_{2}}-6N_{c}\tilde{N}_{c}\\
	&+\frac{1}{4}(1+24n^{2})(2k+N_{f_{1}}+N_{f_{2}})^{2}-3(N_{c}-\tilde{N}_{c}(1-n))(N_{f_{1}}+N_{f_{2}})\\
	&-\frac{3}{2}(2k+N_{f_{1}}+N_{f_{2}})(N_{f_{1}}+N_{f_{2}})(1- n+n^{2})\Big)\\
	&+\frac{3}{4}n m_{A}^{2}\big((N_{f_{1}}+N_{f_{2}})(2k+N_{f_{1}}+N_{f_{2}})-8N_{f_{1}}N_{f_{2}}\big)\\
	&+\frac{3}{4}n\qty((2k+N_{f_{1}}-N_{f_{2}})\sum_{a=1}^{N_{f_{1}}}m_{a}^{2}+(2k-N_{f_{1}}+N_{f_{2}})\sum_{a=1}^{N_{f_{2}}}n_{a}^{2}).
\end{split}
\end{equation}
Again the phase in (\ref{phdav}) can be reproduced from the difference between the contact terms for the global abelian symmetries of the electric and the magnetic theories. We observe here the same unphysical mismatch in $\Delta k_{rr}$ discussed in the $\comm{\mathbf{p}}{\mathbf{p}}_{X,Y}$ case.

\subsection{The $\comm{\mathbf{p}}{\mathbf{0}}_{X,Y}$ case}
The flow to the $\comm{\mathbf{p}}{\mathbf{0}}_{X,Y}$ theory, we start from the $\comm{\mathbf{0}}{\mathbf{0}}_{X,Y}$ $\text{U}(N_{c})_{0}$ duality with $N_{f}$ flavours, and give a positive large real mass to $N_{f}-N_{f_{1}}$ fundamentals. In the IR, this will lead the following duality:
\begin{itemize}
	\item $\text{U}(N_{c})_{k}$ theory with $N_{f_{1}}$ fundamentals and $N_{f}$ anti-fundamentals $Q,\tilde{Q}$, two adjoint $X$ and $Y$ interacting through the superpotential
	\begin{equation}
		\mathcal{W}_{\text{ele}} = \Tr X^{n+1}+\Tr XY^{2}.
	\end{equation}
	\item $\text{U}(\tilde{N}_{c})_{-k}$, where $\tilde{N}_{c}=3nN_{f}-N_{c}$, with $N_{f}$ fundamentals and $N_{f_{1}}$ anti-fundamentals $q,\tilde{q}$, two adjoint fields $x,y$ interacting through the superpotential
	\begin{equation}
	\begin{split}
		\mathcal{W}_{\text{mag}}&=\Tr x^{n+1}+\Tr xy^{2}+\sum_{j=0}^{n-1}\sum_{\ell=0}^{2}\Tr\qty(\mathcal{M}^{j,\ell}qx^{n-1-j}y^{2-\ell}\tilde{q})\\
		&+\sum_{\substack{j=0,\ldots,n-1\\ \ell=0,1,2\\j\ell=0}}V^{+}_{j,\ell}\widetilde{V}^{+}_{n-j,2-\ell}+\sum_{q=0}^{\frac{n-3}{2}}W^{+}_{q}\widetilde{W}^{+}_{\frac{n-3}{2}-q}\;.
	\end{split}
	\end{equation}
\end{itemize}
The CS levels of the two phases are given by $k=\frac{1}{2}(N_{f}-N_{f_{1}})$ and $-k$ respectively. Half of the Coulomb branch is left in this case
and it reflects in the presence of the singlets $V_{j,\ell}^{+}$ and $W_{q}^{+}$ in the spectrum of the dual model.
		
To reproduce this duality on the partition function, we start from the equality (\ref{eq:dualU}) and consider the following shifts for the real masses
\begin{equation}
\begin{cases}
	m_{A}\rightarrow m_{A}+\frac{N_{f}-N_{f_{1}}}{2N_{f}}s\\[3pt]
	m_{a}\rightarrow m_{a}-\frac{N_{f}-N_{f_{1}}}{N_{f}}s& a=1,\ldots,N_{f_{1}}\\
	m_{a}\rightarrow m_{a}+\frac{N_{f_{1}}}{N_{f}}s& a=1,\ldots,N_{f}-N_{f_{1}}\\
	\sigma_{i}\rightarrow \sigma_{i}+\frac{N_{f}-N_{f_{1}}}{2N_{f}}s& i=1,\ldots,N_{c}\\
	\tilde\sigma_{i}\rightarrow \tilde\sigma_{i}+\frac{N_{f}-N_{f_{1}}}{2N_{f}}s& i=1,\ldots,3nN_{f}-N_{c}\\
	\lambda\rightarrow\lambda+(N_{f}-N_{f_{1}})s
\end{cases}
\end{equation}
where we split the axial abelian part, $m_{A}$, for the flavour symmetry from its non-abelian part $m_{a},n_{a}$. 

We study the large $s$ limit as stated before, checking that the leading saddle point contributions cancel between the electric and magnetic partition functions, and we are left with the equality between
\begin{equation}
\begin{split}
	\mathcal{Z}_{\text{ele}}=&\mathcal{Z}_{\text{U}(N_{c})_{k}}^{(N_{f_{1}},N_{f})}(\mu_{a};\nu_{a};\tau_{X};\tau_{Y};\hat\lambda)
\end{split}
\end{equation}
where
\begin{equation}
	\hat\lambda=\lambda+(N_{f}-N_{f_{1}})(m_{A}-\omega)
\end{equation}
and 
\begin{equation}
\begin{split}
	\mathcal{Z}_{\text{mag}} &= e^{i\pi\phi}e^{-\frac{1}{4}\lambda(\frac{3}{2}n\lambda-\eta)}\mathcal{Z}_{\text{U}(\tilde{N}_{c})_{-k}}^{(N_{f},N_{f_{1}})}(\tau_{X}-\tau_{Y}-\nu_{a};\tau_{X}-\tau_{Y}-\mu_{a};\tau_{X};\tau_{Y};\tilde\lambda)\\
	&\times\prod_{j=0}^{n-1}\prod_{	\ell=0}^{2}\prod_{a=1}^{N_{f_{1}}}\prod_{b=1}^{N_{f}}\Gamma_{h}(j\tau_{X}+\ell\tau_{Y}+\mu_{a}+\nu_{b})\\
	&\times\prod_{\substack{j=0,\ldots n-1\\\ell=0,1,2\\j\ell=0}}\Gamma_{h}\qty(\frac{\lambda}{2}+N_{f}\omega-\frac{N_{c}-1}{2}\tau_{X}-\frac{1}{2}\sum_{a=1}^{N_{f}}(m_{a}+n_{a})+j\tau_{X}+\ell\tau_{Y})\\
	&\times\prod_{q=0}^{\frac{n-3}{2}}\Gamma_{h}\qty(\lambda+2N_{f}\omega-(N_{c}-1)\tau_{X}-\sum_{a=1}^{N_{f}}(m_{a}+n_{a})+(2q+1)\tau_{A})
\end{split}
\end{equation}
where $\mu_{a}=m_{a}+m_{A}$ and $\nu_{b}=n_{b}+m_{A}$, which solve the constrain $\sum_{a=1}^{N_{f}}\mu_{a}=\sum_{b=1}^{N_{f}}\nu_{b}=N_{f} m_{A}$, and
\begin{align}
	\tilde\lambda &= -\lambda+(N_{f}-N_{f_{1}})(m_{A}-\tau_{X}+\tau_{Y}+\omega),\\
	\eta &= \tau_{X}+6\tau_{Y}-2\omega+n\Big(6(2k+N_{f_{1}})(\omega-m_{A})+\tau_{X}(2n-3N_{c})-4\omega\Big).
\end{align}
The complex exponent $\phi$ necessary for the equality between the partition functions to hold has the following form
\begin{equation}
\begin{split}
	\phi&=3 m_{A}\tau_{Y}((2k+N_{f_{1}})^{2}(n-2)+3N_{f_{1}}(2k+N_{f_{1}})-N_{c}N_{f_{1}})\\
	&-\frac{\tau_{X}^{2}}{16}-\frac{\tau_{X}\tau_{Y}}{8}\Big((1+n+n^{2})+6N_{c}^{2}+2(\tilde{N}_{c}+N_{c})^{2}-20k(2k+N_{f_{1}})\\
	&+6N_{c}N_{f_{1}}(n-2)+6N_{f_{1}}(2k+N_{f_{1}})(1-2n^{2})+6n(2k+N_{f_{1}})(4k+2N_{f_{1}}-3N_{c})\Big)\\
	&+\frac{3}{2}m_{A}^{2}n(8k^{2}+2kN_{f_{1}}-N_{f_{1}}^{2})+3kn\sum_{a=1}^{N_{f}}n_{a}^{s}\;.
\end{split}
\end{equation}
Again the phase in (\ref{phdav}) can be reproduced from the difference between the contact terms for the global abelian symmetries of the electric and the magnetic theories. We observe here the same unphysical mismatch in $\Delta k_{rr}$ discussed in the $\comm{\mathbf{p}}{\mathbf{p}}_{X,Y}$  and in the $\comm{\mathbf{p}}{\mathbf{q}}_{X,Y}$ case.

\subsection{The $\comm{\mathbf{p}}{\mathbf{q}}_{X,Y}^{*}$ case}
The flow to the $\comm{\mathbf{p}}{\mathbf{q}}^{*}_{X,Y}$ theory, we start from the $\comm{\mathbf{0}}{\mathbf{0}}_{X,Y}$ $\text{U}(N_{c})_{0}$ duality with $N_{f}$ flavours and give a positive large real mass to $N_{f_{1}}$ anti-fundamentals and a negative large real mass to $N_{f_{2}}$ anti-fundamentals. In the IR, this will lead the following duality:
\begin{itemize}
	\item $\text{U}(N_{c})_{k}$ theory with $N_{f}$ fundamentals and $N_{a}=N_{f}-N_{f_{1}}-N_{f_{2}}$ anti-fundamentals $Q,\tilde{Q}$, two adjoint $X$ and $Y$ interacting through the superpotential
	\begin{equation}
		\mathcal{W}_{\text{ele}} = \Tr X^{n+1}+\Tr XY^{2}.
	\end{equation}
	\item $\text{U}(\tilde{N}_{c})_{-k}$, where $\tilde{N}_{c}=3nN_{f}-N_{c}$, with $N_{a}$ fundamentals and $N_{f}$ anti-fundamentals $q,\tilde{q}$, two adjoint fields $x,y$ interacting through the superpotential
	\begin{equation}
	\begin{split}
		\mathcal{W}_{\text{mag}}&=\Tr x^{n+1}+\Tr xy^{2}+\sum_{j=0}^{n-1}\sum_{\ell=0}^{2}\Tr\qty(\mathcal{M}^{j,\ell}qx^{n-1-j}y^{2-\ell}\tilde{q}).\\
	\end{split}
	\end{equation}
\end{itemize}
The CS levels of the two phases are given by $k=\frac{1}{2}(N_{f_{1}}-N_{f_{2}})$ and $-k$ respectively. The CB is lifted and the monopoles acting as singlets in the magnetic theory are integrated out.
	
To reproduce the duality on the  partition function, we start from equality (\ref{eq:dualU}) and consider the following shifts in the real masses
\begin{equation}
\begin{cases}
	m_{A}\rightarrow m_{A}+\frac{N_{f_{1}}-N_{f_{1}}}{2N_{f}}s\\[3pt]
	n_{a}\rightarrow n_{a}-\frac{N_{f_{1}}-N_{f_{2}}}{N_{f}}s& a=1,\ldots,N_{f}-N_{f_{1}}-N_{f_{2}}\\
	n_{a}\rightarrow n_{a}+\frac{N_{f}-N_{f_{1}}+N_{f_{2}}}{N_{f}}s& a=1,\ldots,N_{f_{1}}\\
	n_{a}\rightarrow n_{a}-\frac{N_{f}+N_{f_{1}}-N_{f_{2}}}{N_{f}}s& a=1,\ldots,N_{f_{2}}\\
	\sigma_{i}\rightarrow \sigma_{i}+\frac{N_{f_{2}}-N_{f_{1}}}{2N_{f}}s& i=1,\ldots,N_{c}\\
	\tilde\sigma_{i}\rightarrow \tilde\sigma_{i}+\frac{N_{f_{2}}-N_{f_{1}}}{2N_{f}}s& i=1,\ldots,3nN_{f}-N_{c}\\
	\lambda\rightarrow\lambda-(N_{f_{1}}+N_{f_{2}})s
\end{cases}
\end{equation}
where we split the abelian axial part $m_{A}$ of the real masses for the flavour symmetry from its non abelian part $m_{a},n_{a}$.

We study the large $s$ limit by making use of the asymptotic behavior of the hyperbolic Gamma function (\ref{eq:asymptGamma}). We check that the leading saddle point contributions cancel between the electric and magnetic partition functions, and we are left with the equality between
\begin{equation}
\begin{split}
	\mathcal{Z}_{\text{ele}}=&\mathcal{Z}_{\text{U}(N_{c})_{k}}^{(N_{f},N_{a})}(\mu_{a};\nu_{a};\tau_{X};\tau_{Y};\hat\lambda)
\end{split}
\end{equation}
where
\begin{equation}
	\hat\lambda = \lambda+(N_{f_{1}}-N_{f_{2}})(m_{A}-\omega),
\end{equation}
and 
\begin{equation}
\begin{split}
	\mathcal{Z}_{\text{mag}} &= e^{i\pi\phi}e^{i\pi\lambda\qty(m_{A}(N_{c}+\tilde{N}_{c})-3\tau_{Y}(N_{f}(n+1)-N_{c}))}\mathcal{Z}_{\text{U}(\tilde{N}_{c})_{-k}}^{(N_{f},N_{f_{1}})}(\tau_{X}-\tau_{Y}-\nu_{a};\tau_{X}-\tau_{Y}-\mu_{a};\tilde\lambda)\\
	&\times\prod_{j=0}^{n-1}\prod_{	\ell=0}^{2}\prod_{a=1}^{N_{f}}\prod_{b=1}^{N_{a}}\Gamma_{h}(j\tau_{X}+\ell\tau_{Y}+\mu_{a}+\nu_{b})
\end{split}
\end{equation}
where 
\begin{equation}
	\tilde\lambda=-\lambda-(N_{f_{1}}-N_{f_{2}})(m_{A}-\tau_{X}+\tau_{Y}+\omega),
\end{equation}
with $\mu_{a}=m_{a}+m_{A}$ and $\nu_{b}=n_{b}+m_{A}$ solving the constrain $\sum_{a=1}^{N_{f}}\mu_{a}=\sum_{b=1}^{N_{f}}\nu_{b}=N_{f} m_{A}$.\\
The complex exponent $\phi$ necessary for the equality between the partition functions to hold has the following form
\begin{equation}
\begin{split}
	\phi&= (N_{f_{1}}-N_{f_{2}})\Big(3m_{A}\tau_{Y}(N_{c}-3N_{f})+\frac{\tau_{X}\tau_{Y}}{4}(3N_{c}(n-2)+N_{f}(8-6n^{2}))\\
	&+\frac{9}{2}m_{A}^{2}nN_{f}+\frac{3}{2}n\sum_{a=1}^{N_{f}}m_{a}^{2}\Big).
\end{split}
\end{equation}
In this case the CS contact terms can be computed using the reduction of the $\comm{\mathbf{p}}{\mathbf{q}}^*_{X,Y}$ 
duality to the $\comm{\mathbf{0}}{\mathbf{0}}_{X,Y}$ one.
This requires a Higgsing in the magnetic phase and we obtain the same results discussed above. We match all the contributions except the one of the $\Delta k_{rr}$, but this mismatch is unphysical because it involves a pure phase.

\section{Dualities for $\text{SU}(N_{c})$ chiral SQCD with two adjoints}

As  discussed in section \ref{sec:partFun}, one can start from the $\comm{\mathbf{0}}{\mathbf{0}}$ $\text{U}(N_{c})_{0}$ duality and obtain a duality for $\text{SU}(N_{c})_{0}$ by gauging the topological $\text{U}(1)_{J}$ symmetry \cite{Aharony:2013dha}. This is achieved by introducing a dynamical background multiplet for the topological symmetry. This procedure introduces a mixed CS term in the action
\begin{equation}
	\mathcal{L}\supset A^{\text{U}(1)}\wedge \mathrm{d}\!\Tr A^{\text{U}(N_{c})}
\end{equation}
at level $-1$  between the new $\text{U}(1)$ symmetry coming from the topological $\text{U}(1)_{J}$ and the abelian subgroup of the gauge symmetry $\text{U}(1)\subset \text{U}(N_{c})$. In addition a new topological $\text{U}(1)_{J^{\prime}}$ is generated from the hodge dual of the gauged $\text{U}(1)_{J}$ field strength which is conserved by virtue of the Bianchi identity. In the absence of monopoles, which are charged under the gauged topological symmetry, the mixed CS term makes the two $\text{U}(1)$ photons massive and can be integrated out in the IR. In this case, the gauge group becomes $\text{SU}(N_{c})_{0}$ and the topological $\text{U}(1)_{J^{\prime}}$ can be considered as the baryonic symmetry $\text{U}(1)_{B}$ under which the flavour has canonically normalized charge $1/N_{c}$.\\
In  presence of fields charged under  $\text{U}(1)_{J}$ the analysis has to be modified. It is the case for example of
 many of the dual phases, where the electric monopoles are singlets in the dual description.
It follows that we cannot  decouple the dynamics of the gauged $\text{U}(1)_{J}$  and of the $\text{U}(1)\subset \text{U}(\tilde{N}_{c})$ symmetries. In this case, the magnetic side will be a $\text{U}(\tilde{N}_{c})_{0}\times \text{U}(1)$ gauge theory with a level $-1$ mixed CS term.
In some case some further local duality simplifies such sectors. For example when considering  Aharony duality one can use the SQED/XYZ duality.

Another common feature of the  gauging of the topological symmetry leading from $\text{U}(N_{c})$ to $\text{SU}(N_{c})$ in  presence of adjoint matter, consists of  imposing the tracelessness condition \cite{Amariti:2020xqm,Benvenuti:2020gvy} on the adjoints. 
This can be achieved in two ways: one can either add a flipping term in the superpotential on the magnetic side
\begin{equation}
	\mathcal{W}_{\text{flip}} = \alpha_{0}\Tr x+\beta_{0}\Tr y,
\end{equation}
which imposes then the tracelessness of the adjoints by the F-term equations for the singlets $\alpha_{0}$ and $\beta_{0}$, or consider the traceless adjoint representation of $\text{U}(\tilde{N}_{c})$.

The procedure just described gives the following duality
\begin{itemize}
\item $3d$ $\mathcal{N}=2$ $\text{SU}(N_{c})_{0}$ SQCD with $N_{f}$ flavours $Q,\tilde{Q}$ and two adjoint fields $X,Y$, with superpotential
	\begin{equation}
		\mathcal{W}_{\text{ele}}=\Tr X^{n+1}+\Tr XY^{2}.
	\end{equation}
\item $3d$ $\mathcal{N}=2$ $\text{U}(\tilde{N}_{c})_{0}\times \text{U}(1)$ SQCD with $\tilde{N}_{c}=3nN_{f}-N_{c}$, $N_{f}$ dual flavours $q,\tilde{q}$ in the non-abelian sector and two adjoints $X,Y$, $2(n-1)$ pairs of fields $V^{\pm}_{j,\ell}$ and $\frac{1}{2}(n-1)$ pairs of  fields $W^{\pm}_{q}$ in the abelian gauge sector with opposite gauge charge. 
These fields interact with the dressed monopoles and anti-monopoles of the 
$\text{U}(\tilde{N}_{c})$ sector, that in this case carry $\pm 1$ charge under the new $\text{U}(1)$ gauged factor as well.
There is also a level $-1$ CS level between the abelian $\text{U}(1)$ subgroup in the $\text{U}(\tilde{N}_{c})$ and the other $\text{U}(1)$ gauge factor. The superpotential is given by
	\begin{equation}
	\begin{split}
		\mathcal{W}_{\text{mag}}&=\Tr x^{n+1}+\Tr xy^{2}+\alpha_{0}\Tr x+\beta_{0}\Tr y+\Tr \sum_{j=0}^{n-1}\sum_{\ell=0}^{2}\Tr\qty(\mathcal{M}^{j,\ell}qx^{n-1-j}y^{2-\ell}\tilde{q})\\
		&+\sum_{\substack{j=0,\ldots,n-1\\\ell=0,1,2\\j\ell=0}}V^{\pm}_{j,\ell}\widetilde{V}^{\pm}_{n-j,2-\ell}+\sum_{q=0}^{\frac{n-3}{2}}W^{\pm}_{q}\widetilde{W}^{\pm}_{\frac{n-3}{2}-q}\;.
	\end{split}
	\end{equation}
\end{itemize}
The non-anomalous global symmetry of the theories is $\text{SU}(N_{f})_{L}\times\text{SU}(N_{f})_{R}\times \text{U}(1)_{A}\times\text{U}(1)_{B}\times\text{U}(1)_{R}$ under which the fields are charged as in table \ref{tab:SU}.\\

\begin{table}
\centering
\caption{Matter content of the electric (upper) and magnetic (lower) theories after gauging the topological symmetry. The subscript shows the charge of the fields under the gauged $\text{U}(1)_{J}$.}
\vspace{2pt}
\renewcommand{\arraystretch}{1.2}
\adjustbox{max width=\columnwidth}{
\begin{tabular}{||c|c|c|c|c|c|c||}
	\cline{2-7}
	\multicolumn{1}{c|}{}&Gauge&\multicolumn{5}{c||}{Global}\\
	\hline
	Field & SU$(N_{c})_{0}$ & SU$(N_{f})_{L}$ & SU$(N_{f})_{R}$ & U$(1)_{A}$ & U$(1)_{B}$ & U$(1)_{R}$\\
	\hline
	$Q$ & $\ydiagram{1}$ & $\ydiagram{1}$ & $1$ & $1$ & $1$ & $r_{Q}$\\
	$\tilde Q$ & $\overline{\ydiagram{1}}$ & $1$ & $\overline{\ydiagram{1}}$ & $1$ & $-1$ & $r_{Q}$\\
	$X$ & Adj & $1$ & $1$ & $0$ & $0$ & $\frac{2}{n+1}$\\
	$Y$ & Adj & $1$ & $1$ & $0$ & $0$ & $\frac{n}{n+1}$\\[4pt]
	\hline\hline
	Field & U$(\tilde N_{c})_{0}\times$U$(1)$ & SU$(N_{f})_{L}$ & SU$(N_{f})_{R}$ & U$(1)_{A}$ & U$(1)_{B}$ & U$(1)_{R}$\\
	\hline
	$q$ & $\overline{\ydiagram{1}}_{\,0}$ & $1$ & $\overline{\ydiagram{1}}$ & $-1$ & $0$ & $\frac{2-n}{n+1}-r_{Q}$\\
	$\tilde q$ & $\ydiagram{1}_{\,0}$ & $\ydiagram{1}$ & $1$ & $-1$ & $0$ & $\frac{2-n}{n+1}-r_{Q}$\\
	$x$ & $\text{Adj}_{\,0}$ & $1$ & $1$ & $0$ & $0$ & $\frac{2}{n+1}$\\
	$y$ & $\text{Adj}_{\,0}$ & $1$ & $1$ & $0$ & $0$ & $\frac{n}{n+1}$\\[4pt]
	$\mathcal{M}_{j\ell}$ & $1_{0}$ & $\Box$ & $\overline\Box$ & $2$ & $0$ & $2r_{Q}+\frac{2j+n\ell}{n+1}$\\
	$V_{j\ell}^{\pm}$ & $1_{\pm 1}$ & $1$ & $1$ & $-N_{f}$ & $\pm N_{c}$ &$(1-r_{Q})N_{f}+\frac{2j+n\ell-(N_{c}-1)}{n+1}$\\
	$W_{q}^{\pm}$ & $1_{\pm 2}$ & $1$ & $1$ & $-2N_{f}$ & $\pm N_{c}$ & $2(1-r_{Q})N_{f}+\frac{2+4q-2(N_{c}-1)}{n+1}$\\
	\hline
\end{tabular}}
\label{tab:SU}
\end{table}

At the level of the partition function, the gauging procedure, is implemented by adding a factor $\frac{1}{2}e^{i\pi\lambda N_{c}m_{B}}$ to both side of the identity (\ref{eq:dualU}) and then integrating over $\lambda$. The integral over the FI corresponds to the gauging of the $\text{U}(1)_{J}$ while the added exponential factor carries the additional baryonic symmetry, whose real mass we label as $m_{B}$. The numerical factor $1/2$ is added to ensure the proper normalization of the $V_{j,\ell}^{\pm}$ and $W^{\pm}_{q}$ under the gauged $\text{U}(1)_{J}$. On the electric side, the only dependence on the FI is in the exponential term that can be integrated upon shifting the Cartan variables by $m_{B}$. This will lead to flavour matter fields carrying baryonic charge. On the magnetic side, the
fields $V^{\pm}_{j,\ell}$ and  $W^{\pm}_{q}$
 are charged under the topological $U(1)_{J}$ and therefore the integration is not straightforward, i.e. a local mirror duality should be necessary to get rid of this sector. 
  For our purpose we will leave the integration on $\lambda$ explicit on both sides whilst shifting the Cartan on the electric side, making explicit the baryonic symmetry of the matter. This will lead us to the identity between the electric partition function
\begin{equation}
\begin{split}
	\mathcal{Z}_{\text{ele}} &= \frac{\Gamma_{h}(\tau_{X})^{N_{c}-1}\Gamma_{h}(\tau_{Y})^{N_{c}-1}}{N_{c}!\sqrt{-\omega_{1}\omega_{2}}^{N_{c}}}\int\dd{\xi}\int\prod_{i=1}^{N_{c}}\dd{\sigma_{i}}\exp(-2i\pi\xi\sigma_{i})\\
	&\times\prod_{i=1}^{N_{c}}\prod_{a=1}^{N_{f}}\Gamma_{h}(\mu_{a}+m_{B}+\sigma_{i};\nu_{a}-m_{B}-\sigma_{i})\prod_{1\le i<j\le N_{c}}\prod_{\beta=X,Y}\frac{\Gamma_{h}(\tau_{\beta}\pm(\sigma_{i}-\sigma_{j}))}{\Gamma_{h}(\pm(\sigma_{i}-\sigma_{j}))}
\end{split}
\label{eq:SUele}
\end{equation}
and the magnetic partition function
\begin{equation}
\begin{split}
	\mathcal{Z}_{\text{mag}}&=\frac{\Gamma_{h}(\tau_{X})^{\tilde{N}_{c}-1}\Gamma_{h}(\tau_{Y})^{\tilde{N}_{c}-1}}{\tilde{N_{c}}!\sqrt{-\omega_{1}\omega_{2}}^{\tilde{N}_{c}}}\prod_{j=0}^{n-1}\prod_{\ell=0}^{2}\prod_{a,b=1}^{N_{f}}\Gamma_{h}(j\tau_{X}+\ell\tau_{Y}+\mu_{a}+\nu_{b})\\
	&\times\int\dd{\xi}\int\prod_{i=1}^{\tilde{N}_{c}}\dd{\sigma_{i}}\exp(2i\pi\xi\qty(\sigma_{i}+\frac{N_{c}}{\tilde{N}_{c}}m_{B}))\prod_{a=1}^{N_{f}}\Gamma_{h}(\tau_{X}-\tau_{Y}-\nu_{a}+\sigma_{i};\tau_{X}-\tau_{Y}-\mu_{a}-\sigma_{i})\\
	&\times\prod_{1\le i<j\le \tilde{N}_{c}}\prod_{\beta=X,Y}\frac{\Gamma_{h}(\tau_{\beta}\pm(\sigma_{i}-\sigma_{j}))}{\Gamma_{h}(\pm(\sigma_{i}-\sigma_{j}))}\\
	&\times\prod_{\substack{j=0,\ldots,n-1\\\ell=0,1,2\\j\ell=0}}\Gamma_{h}\qty(\pm\xi+N_{f}\omega-\frac{N_{c}-1}{2}\tau_{X}-\frac{1}{2}\sum_{a=1}^{N_{f}}(\mu_{a}+\nu_{a})+j\tau_{X}+\ell\tau_{Y})\\
	&\times\prod_{q=0}^{\frac{n-3}{2}}\Gamma_{h}\qty(\pm2\xi+2N_{f}\omega+(N_{c}-1)\tau_{X}-\sum_{a=1}^{N_{f}}(\mu_{a}+\nu_{a})+(2q+1)\tau_{X})\\
\end{split}
\label{eq:SUmag}
\end{equation}
where we rescaled the FI as $\lambda=2\xi$ \footnote{With this normalization, the exponential term for the FI has an added factor $2$ which will be implicit.}. The addition of the flipping terms in the magnetic superpotential is reflected on the partition function by an additional $\Gamma_{h}(2\omega-\tau_{X};2\omega-\tau_{Y})$ factor. Using the inversion formula (\ref{eq:inversion}), these factors cancel out with a factor of $\Gamma_{h}(\tau_{X})$ and $\Gamma_{h}(\tau_{Y})$.\\

\subsection{The $\comm{\mathbf{p}}{\mathbf{p}}_{X,Y}$ case}
To flow to the $\comm{\mathbf{p}}{\mathbf{p}}_{X,Y}$ theory, we start from the $\comm{\mathbf{0}}{\mathbf{0}}_{X,Y}$ $\text{SU}(N_{c})_{0}$ duality with $N_{f}+k$ fundamentals and anti-fundamentals and give a positive large, but finite, real mass to $k$ flavours. In the IR, this will lead to the following duality
\begin{itemize}
	\item $\text{SU}(N_{c})_{k}$ SQCD with $N_{f}$ fundamentals and anti-fundamentals $Q,\tilde{Q}$, two adjoints $X,Y$ interacting through the superpotential
	\begin{equation}
		\mathcal{W}_{\text{ele}}=\Tr X^{n+1}+\Tr X Y^{2}.
	\end{equation}
	\item $\text{U}(\tilde{N}_{c})_{-k}\times\text{U}(1)_{3n}$ SQCD with $\tilde{N}_{c}=3n(N_{f}+\abs{k})-N_{c}$,  a level $-1$ mixed CS, $N_{f}$ dual fundamentals and anti-fundamentals $q,\tilde{q}$, two traceless adjoint fields $x,y$ interacting through the superpotential
	\begin{equation}
		\mathcal{W}_{\text{mag}}=\Tr x^{n+1}+\Tr xy^{2}+ \sum_{j=0}^{n-1}\sum_{\ell=0}^{2}\Tr\qty(\mathcal{M}^{j,\ell}qx^{n-1-j}y^{2-\ell}\tilde{q}).
	\end{equation}
\end{itemize}
 The fields $V^{\pm}_{j,\ell}$ and  $W^{\pm}_{q}$ in the dual phase are  massive and they are integrated out. The electric and the magnetic theories acquire a CS level $k$ and $-k$ respectively. From here on we omit in the magnetic phase the singlets needed to make the adjoint traceless since we can integrate them out in the IR.\\
At this stage, one could also decouple the dynamics of the massive photons in the two gauge $\text{U}(1)$ in the magnetic side, because we do not have matter charged under $\text{U}(1)_{J}$. 

On the partition function, the real mass flow that produces the above-mentioned duality, is given by the assignment of real masses of (\ref{eq:shiftUpp}). The real mass associated to the baryonic $\text{U}(1)_{B}$ symmetry does not get shifted.\\
We study the limit of large $s$ on both the electric (\ref{eq:SUele}) and magnetic (\ref{eq:SUmag}) partition functions by making use of the asymptotic behavior of the hyperbolic Gamma function (\ref{eq:asymptGamma}). We check that the leading saddle point contributions cancel between the electric and magnetic phases, and we are left with the equality between
\begin{equation}
	\mathcal{Z}_{\text{ele}} = \mathcal{Z}_{\text{SU}(N_{c})_{k}}^{(N_{f},N_{f})}(\mu_{a};\nu_{a};\tau_{X};\tau_{Y})
\end{equation}
in which we have no FI since after the flow we integrate on $\xi$, and
\begin{equation}
\begin{split}
	\mathcal{Z}_{\text{mag}} &=  \prod_{j=0}^{n-1}\prod_{\ell=0}^{2}\prod_{a,b=1}^{N_{f}}\Gamma_{h}(j\tau_{X}+\ell\tau_{Y}+\mu_{a}+\nu_{b})\\
	&\times e^{i\pi\phi}\int\dd{\xi}e^{i\pi(2m_{B}N_{c}\xi-3n\xi^{2})}\mathcal{Z}_{\text{U}(\tilde{N}_{c})_{-k}}^{(N_{f},N_{f})}(\tau_{X}-\tau_{Y}-\nu_{a}; \tau_{X}-\tau_{Y}-\mu_{a};\tau_{X};\tau_{Y};-\xi).
	\label{eq:SUmagpp}
\end{split}
\end{equation}
From (\ref{eq:SUmagpp}) we can see the level $3n$ CS of the $\text{U}(1)$ gauge factor and the $-1$ mixed CS, in the term $m_{B}\xi$, between the abelian subgroup of the $\text{U}(\tilde{N}_{c})$ and the abelian $\text{U}(1)$ gauge group. Again the real masses satisfy the constrain $\sum_{a=1}^{N_{f}}\mu_{a}=\sum_{a=1}^{N_{f}}\nu_{a}=N_{f}m_{A}$ where $\mu_{a}=m_{a}+m_{A}$ and $\nu_{a}=n_{a}+m_{A}$.

The complex exponent $\phi$ needed for the matching is given by
\begin{equation}
\begin{split}
	\phi&= 3m_{A}^{2}nN_{f}(k-N_{f})-k m_{B}^{2}N_{c}+\frac{\tau_{X}}{2}\omega m_{A}\Big(3n(n+1)(2N_{f}^{2}+k^{2}-(2n^{2}+1)(N_{f}+k)\\
	&-3N_{c}n(N_{f}-k)+kN_{f}(6n(n-1)+2N_{c})\Big)-\frac{\tau_{X}^{2}}{8}\Big(1+N_{c}(2k+1)-2(N_{f}+k)\Big)\\
	&-\frac{\tau_{X}\tau_{Y}}{8}\Big((2n^{2}+8n-1)-4(N_{f}+k)(1+2n+2n^{2})+12(1+2n+n^{2})N_{f}^{2}\\
	&+(54n^{2}+33n-1)k^{2}+2(30n^{2}+33n-1)N_{f}k-12(n+1)N_{f}N_{c}-4(6n+1)kN_{c}\\
	&+N_{c}(3N_{c}+4n)\Big)+\frac{3}{2}kn\sum_{a=1}^{N_{f}}\qty(m_{a}^{2}+n_{a}^{2}).
\end{split}
\end{equation}
\subsection{The $\comm{\mathbf{p}}{\mathbf{q}}_{X,Y}$ case}
To flow to the $\comm{\mathbf{p}}{\mathbf{q}}_{X,Y}$ theory, we start from the $\comm{\mathbf{0}}{\mathbf{0}}_{X,Y}$ $\text{SU}(N_{c})_{0}$ duality with $N_{f}$ fundamentals and anti-fundamentals and give a positive large real mass to $N_{f}-N_{f_{1}}$ fundamentals and $N_{f}-N_{f_{2}}$ anti-fundamentals. This will lead to the following duality
\begin{itemize}
	\item $\text{SU}(N_{c})_{k}$ SQCD with $N_{f_{1}}$ fundamentals and $N_{f_{2}}$ anti-fundamentals $Q,\tilde{Q}$, two adjoints $X,Y$ interacting through the superpotential
	\begin{equation}
		\mathcal{W}_{\text{ele}}=\Tr X^{n+1}+\Tr X Y^{2}.
	\end{equation}
	\item $\text{U}(\tilde{N}_{c})_{-k}\times\text{U}(1)_{3n}$ SQCD with $\tilde{N}_{c}=3nN_{f}-N_{c}$, a level $-1$ mixed CS between the two $\text{U}(1)s$, $N_{f_{2}}$ dual fundamentals and $N_{f_{1}}$ dual anti-fundamentals $q,\tilde{q}$, two traceless adjoint fields $x,y$ interacting through the superpotential
	\begin{equation}
		\mathcal{W}_{\text{mag}}=\Tr x^{n+1}+\Tr xy^{2}+ \sum_{j=0}^{n-1}\sum_{\ell=0}^{2}\Tr\qty(\mathcal{M}^{j,\ell}qx^{n-1-j}y^{2-\ell}\tilde{q}).
	\end{equation}
\end{itemize}
 The fields $V^{\pm}_{j,\ell}$ and  $W^{\pm}_{q}$ in the dual phase are  massive and they are integrated out. The electric and the magnetic theories acquire a CS level $k=N_{f}-\frac{1}{2}\qty(N_{f_{1}}+N_{f_{2}})$ and $-k$ respectively.

To reproduce this duality at the level of the partition function, we start from the equality between (\ref{eq:SUele}) and (\ref{eq:SUmag}) and consider the following shifts in the real masses
\begin{equation}
\begin{cases}
	m_{A}\rightarrow m_{A}+\frac{2N_{f}-N_{f_{1}}-N_{f_{2}}}{2N_{f}}s\\
	m_{B}\rightarrow m_{B}-\frac{N_{f_{1}}-N_{f_{2}}}{2N_{f}}s\\
	m_{a}\rightarrow m_{a}-\frac{N_{f}-N_{f_{1}}}{N_{f}}s&a=1,\ldots,N_{f_{1}}\\
	m_{a}\rightarrow m_{a}+\frac{N_{f_{1}}}{N_{f}}s&a=1,\ldots,N_{f}-N_{f_{1}}\\
	n_{a}\rightarrow n_{a}-\frac{N_{f}-N_{f_{2}}}{N_{f}}s&a=1,\ldots,N_{f_{2}}\\
	n_{a}\rightarrow n_{a}+\frac{N_{f_{2}}}{N_{f}}s&a=1,\ldots,N_{f}-N_{f_{2}}\\
	\tilde{\sigma}_{i}\rightarrow\tilde{\sigma}_{i}-\frac{N_{f_{1}}-N_{f_{2}}}{2N_{f}}s&i=1,\ldots,3nN_{f}-N_{c}\\
	\xi\rightarrow\xi-\frac{N_{f_{1}}-N_{f_{2}}}{2}s
\end{cases}
\end{equation}
We study the limit of large $s$ on both the electric (\ref{eq:SUele}) and magnetic (\ref{eq:SUmag}) partition functions by making use of the asymptotic behavior of the hyperbolic Gamma function (\ref{eq:asymptGamma}). We check that the leading saddle point contributions cancel between the electric and magnetic phases, and we are left with the equality between
\begin{equation}
	\mathcal{Z}_{\text{ele}} = \mathcal{Z}_{\text{SU}(N_{c})_{k}}^{(N_{f_{1}},N_{f_{2}})}(\mu_{a};\nu_{a};\tau_{X};\tau_{Y})
\end{equation}
in which we have no FI since after the flow we integrate on $\xi$, and
\begin{equation}
\begin{split}
	\mathcal{Z}_{\text{mag}} &=  \prod_{j=0}^{n-1}\prod_{\ell=0}^{2}\prod_{a=1}^{N_{f_{1}}}\prod_{b=1}^{N_{f_{2}}}\Gamma_{h}(j\tau_{X}+\ell\tau_{Y}+\mu_{a}+\nu_{b})\\
	&\times e^{i\pi\phi}\int\dd{\xi}e^{i\pi(2m_{B}N_{c}\xi-3n\xi^{2})}\mathcal{Z}_{\text{U}(\tilde{N}_{c})_{-k}}^{(N_{f_{1}},N_{f_{2}})}(\tau_{X}-\tau_{Y}-\nu_{a}; \tau_{X}-\tau_{Y}-\mu_{a};\tau_{X};\tau_{Y};\hat{\xi})
	\label{eq:SUmagpq}
\end{split}
\end{equation}
where
\begin{equation}
	\hat{\xi}=\xi-\frac{N_{f_{1}}-N_{f_{2}}}{2}(m_{A}-\tau_{X}+\tau_{Y}+\omega).
\end{equation}
From (\ref{eq:SUmagpq}) we can see the level $3n$ CS of the $\text{U}(1)$ gauge factor and the $-1$ mixed CS, in the term $m_{B}\xi$, between the abelian subgroup of the $\text{U}(\tilde{N}_{c})$ and the abelian $\text{U}(1)$ gauge group. Again the real masses satisfy the constrain $\sum_{a=1}^{N_{f}}\mu_{a}=\sum_{a=1}^{N_{f}}\nu_{a}=N_{f}m_{A}$ where $\mu_{a}=m_{a}+m_{A}$ and $\nu_{a}=n_{a}+m_{A}$.

The complex exponent $\phi$ needed for the matching is given by
\begin{equation}
\begin{split}
	\phi&= 3m_{A}^{2} n \Big(N_{f}^{2}-N_{f}(4N_{f_{2}}+k)+2N_{f_{2}}(N_{f_{2}}+2k))\Big)-m_{B}^{2}kN_{c}\\
	&+\frac{\tau_{X}}{2}m_{A}\Big(2nN_{f_{2}}^{2}(n-3)+3nN_{f}^{2}(3n-1)+2k(N_{c}(1+3n)+3nN_{f}(n-3))\\
	&+N_{f}(2+2n^{2}+12kn^{2}+3nN_{c}+6nN_{f_{2}}(n-3))\Big)+2m_{B}\omega N_{c}(N_{f_{2}}-N_{f}+k)\\
	&-\frac{\tau_{X}^{2}}{8}\Big(1+2N_{f}+N_{c}(1+2k)\Big)-\frac{\tau_{X}\tau_{Y}}{8}\Big(-1+8n+2n^{2}+N_{f}^{2}(-1+33n+6n^{2})\\
	&+2N_{f_{2}}k(-13+9n-6n^{2})+N_{f_{2}}(-13+9n-6n^{2})+2N_{f}\big((-2+13N_{f_{2}}-6N_{c})\\
	&-n(4+9N_{f_{2}}+6N_{c})+n^{2}(-4+6N_{f_{2}}+24k)\big)+N_{c}(4n+k(8-12n))\\
	&+3N_{c}^{2}\Big)+\frac{3}{2}n\qty((N_{f}-N_{f_{2}})\sum_{a=1}^{N_{f_{1}}}\mu_{a}^{2}+(N_{f}-N_{f_{1}})\sum_{a=1}^{N_{f_{2}}}\nu_{a}^{2}).
\end{split}
\end{equation}

\subsection{The $\comm{\mathbf{p}}{\mathbf{0}}_{X,Y}$ case}
To flow to the $\comm{\mathbf{p}}{\mathbf{0}}_{X,Y}$ theory, we start from the $\comm{\mathbf{0}}{\mathbf{0}}_{X,Y}$ $\text{SU}(N_{c})_{0}$ duality with $N_{f}$ flavours and give a positive large real mass to $N_{f}-N_{f_{1}}$ fundamentals. This will lead to the following duality
\begin{itemize}
	\item $\text{SU}(N_{c})_{k}$ SQCD with $N_{f_{1}}$ fundamentals and $N_{f}$ anti-fundamentals $Q,\tilde{Q}$, two 	adjoints $X$ and $Y$ interacting through the superpotential
	\begin{equation}
		\mathcal{W}_{\text{ele}}=\Tr X^{n+1}+\Tr XY^{2}.
	\end{equation}
	\item $\text{U}(\tilde{N}_{c})_{-k}\times\text{U}(1)_{\frac{3}{2}n}$ SQCD with $\tilde{N}_{c}=3nN_{f}-N_{c}$,  a level $-1$ mixed CS between the two $\text{U}(1)s$, $N_{f}$ dual fundamentals and $N_{f_{1}}$ dual anti-fundamentals $q,\tilde{q}$, two traceless adjoints $x,y$ interacting through the superpotential
	\begin{equation}
	\begin{split}
		\mathcal{W}_{\text{mag}}&=\Tr x^{n+1}+\Tr xy^{2}+\sum_{j=0}^{n-1}\sum_{\ell=0}^{2}\Tr\qty(\mathcal{M}^{j,\ell}qx^{n-1-j}y^{2-\ell}\tilde q)\\
		&+\sum_{\substack{j=0,\ldots,n-1\\\ell=0,1,2\\j\ell=0}}V^{+}_{j,\ell}\widetilde{V}^{+}_{n-j,2-\ell}+\sum_{q=0}^{\frac{n-3}{2}}W^{+}_{q}\widetilde{W}^{+}_{\frac{n-3}{2}-q}\;.
	\end{split}
	\end{equation}
\end{itemize}
 The fields $V^{\pm}_{j,\ell}$ and  $W^{\pm}_{q}$ in the dual phase are  massive and they are integrated out.  The electric and the magnetic theories acquire a CS level $k=\frac{1}{2}(N_{f}-N_{f_{1}})$ and $-k$ respectively.

To reproduce this duality on the partition function, we start from the equality between (\ref{eq:SUele}) and (\ref{eq:SUmag}) and consider the following shifts of the real masses
\begin{equation}
\begin{cases}
	m_{A}\rightarrow m_{A}+\frac{N_{f}-N_{f_{1}}}{2N_{f}}s\\
	m_{B}\rightarrow m_{B}+\frac{N_{f}-N_{f_{1}}}{2N_{f}}s\\
	m_{a}\rightarrow m_{a}-\frac{N_{f}-N_{f_{1}}}{N_{f}}s&a=1,\ldots,N_{f_{1}}\\
	m_{a}\rightarrow m_{a}+\frac{N_{f_{1}}}{N_{f}}s&a=1,\ldots,N_{f}-N_{f_{1}}\\
	\tilde\sigma_{i}\rightarrow\tilde\sigma_{i}+\frac{N_{f}-N_{f_{1}}}{2N_{f}}s&i=1,\ldots,3nN_{f}-N_{c}\\
	\xi\rightarrow\xi+\frac{N_{f}-N_{f_{1}}}{2}s
\end{cases}
\end{equation}

We study the limit of large $s$ on both the electric (\ref{eq:SUele}) and magnetic (\ref{eq:SUmag}) partition functions by making use of the asymptotic behavior of the hyperbolic Gamma function (\ref{eq:asymptGamma}). We check that the leading saddle point contributions cancel between the electric and magnetic phases, and we are left with the equality between
\begin{equation}
	\mathcal{Z}_{\text{ele}} = \mathcal{Z}_{\text{SU}(N_{c})_{k}}^{(N_{f_{1}},N_{f})}(\mu_{a};\nu_{a};\tau_{X};\tau_{Y})
\end{equation}
in which we have no FI since after the flow we integrate on $\xi$, and
\begin{equation}
\begin{split}
	\mathcal{Z}_{\text{mag}} &=  \prod_{j=0}^{n-1}\prod_{\ell=0}^{2}\prod_{a=1}^{N_{f_{1}}}\prod_{b=1}^{N_{f}}\Gamma_{h}(j\tau_{X}+\ell\tau_{Y}+\mu_{a}+\nu_{b})\\
		&\times e^{i\pi\phi}\int\dd{\xi}e^{i\frac{\pi}{2}\xi(\eta-3n\xi)}\mathcal{Z}_{\text{U}(\tilde{N}_{c})_{-k}}^{(N_{f},N_{f_{1}})}(\tau_{X}-\tau_{Y}-\nu_{a}; \tau_{X}-\tau_{Y}-\mu_{a};\tau_{X};\tau_{Y};\hat\xi)\\
	&\times\prod_{q=0}^{\frac{n-3}{2}}\Gamma_{h}\qty(2\xi+2N_{f}\omega-(N_{c}-1)\tau_{X}-\sum_{a=1}^{N_{f}}(m_{a}+n_{a})+(2q+1)\tau_{A})\\
	&\times\prod_{\substack{j=0,\ldots n-1\\\ell=0,1,2\\j\ell=0}}\Gamma_{h}\qty(\xi+N_{f}\omega-\frac{N_{c}-1}{2}\tau_{X}-\frac{1}{2}\sum_{a=1}^{N_{f}}(m_{a}+n_{a})+j\tau_{X}+\ell\tau_{Y})
\label{eq:SUmagp0}
\end{split}
\end{equation}
where
\begin{equation}
\begin{split}
	\hat\xi &= -\xi+\frac{N_{f}-N_{f_{1}}}{2}(m_{A}-\tau_{X}+\tau_{Y}+\omega),\\
	\eta &=4m_{B}N_{c}-6nm_{A}N_{f}+(1-3nN_{c}+2n^{2})\tau_{X}+6\tau_{Y}-2\omega(1+2n-3nN_{f}).
\end{split}
\end{equation}
From (\ref{eq:SUmagp0}) we can see the level $\frac{3}{2}n$ CS of the $\text{U}(1)$ gauge factor and the $-1$ mixed CS, in the term $m_{B}\xi$, between the abelian subgroup of the $\text{U}(\tilde{N}_{c})$ and the abelian $\text{U}(1)$ gauge group. Again the real masses satisfy the constrain $\sum_{a=1}^{N_{f}}\mu_{a}=\sum_{a=1}^{N_{f}}\nu_{a}=N_{f}m_{A}$ where $\mu_{a}=m_{a}+m_{A}$ and $\nu_{a}=n_{a}+m_{A}$.

The complex exponent $\phi$ needed for the matching is given by
\begin{equation}
\begin{split}
	\phi&=-\frac{3}{2}m_{A}^{2}nN_{f}(N_{f}-6k)-kN_{c}m_{B}^{2}+\frac{\tau_{X}}{4}m_{A}\Big(6nN_{f}^{2}(n+1)+4kN_{c}(1+3n)\\
	&+N_{f}\big(1+2n^{2}+3nN_{c}+12kn(3+n)\big)\Big)+2kN_{c}m_{B}\omega-\frac{\tau_{X}^{2}}{16}\Big(1-2N_{f}+N_{c}(1+4k)\Big)\\
	&-\frac{\tau_{X}\tau_{Y}}{16}\Big((-1+8n+2n^{2})+3N_{c}^{2}+12N_{f}^{2}(1+n)^{2}+4N_{f}\big(-(1+2n+2n^{2})\\
	&-3N_{c}(1+n)+(-13+9n+18n^{2})k\big)+N_{c}\big(4n+k(16-24n)\big)\Big)+3kn\sum_{a=1}^{N_{f}}n_{a}^{2}\;.
\end{split}
\end{equation}

\subsection{The $\comm{\mathbf{p}}{\mathbf{q}}_{X,Y}^{*}$ case}
To flow to the $\comm{\mathbf{p}}{\mathbf{q}}_{X,Y}^{*}$ theory, we start from the $\comm{\mathbf{0}}{\mathbf{0}}_{X,Y}$ $\text{SU}(N_{c})_{0}$ duality with $N_{f}$ flavours and give a positive large real mass to $N_{f_{1}}$ anti-fundamentals and a negative large real mass to $N_{f_{2}}$ anti-fundamentals. This will lead to the following duality
\begin{itemize}
	\item $\text{SU}(N_{c})_{k}$ SQCD with $N_{f}$ fundamentals and $N_{a}=N_{f}-N_{f_{1}}-N_{f_{2}}$ anti-fundamentals $Q,\tilde{Q}$, two 	adjoints $X$ and $Y$ interacting through the superpotential
	\begin{equation}
		\mathcal{W}_{\text{ele}}=\Tr X^{n+1}+\Tr XY^{2}.
	\end{equation}
	\item $\text{U}(\tilde{N}_{c})_{-k}\times\text{U}(1)$ SQCD with $\tilde{N}_{c}=3nN_{f}-N_{c}$,  a level $-1$ mixed CS between the two $\text{U}(1)s$, $N_{a}$ dual fundamentals and $N_{f}$ dual anti-fundamentals $q,\tilde{q}$, two traceless adjoints $x,y$ interacting through the superpotential
	\begin{equation}
		\mathcal{W}_{\text{mag}}=\Tr x^{n+1}+\Tr xy^{2}+\sum_{j=0}^{n-1}\sum_{\ell=0}^{2}\Tr\qty(\mathcal{M}^{j,\ell}qx^{n-1-j}y^{2-\ell}\tilde q).
	\end{equation}
\end{itemize}
 The fields $V^{\pm}_{j,\ell}$ and  $W^{\pm}_{q}$ in the dual phase are  massive and they are integrated out. The
 electric and the magnetic theories acquire a CS level $k=\frac{1}{2}(N_{f_{1}}-N_{f_{2}})$ and $-k$ respectively.

To reproduce this duality on the partition function we start from the equality between (\ref{eq:SUele}) and (\ref{eq:SUmag}) and consider the following shifts of the real masses
\begin{equation}
\begin{cases}
	m_{A}\rightarrow m_{A}+\frac{N_{f_{1}}-N_{f_{2}}}{2N_{f}}s\\
	m_{B}\rightarrow m_{B}-\frac{N_{f_{1}}-N_{f_{2}}}{2N_{f}}s\\
	n_{a}\rightarrow n_{a}-\frac{N_{f_{1}}-N_{f_{2}}}{N_{f}}s&a=1,\ldots,N_{f}-N_{f_{1}}-N_{f_{2}}\\
	n_{a}\rightarrow n_{a}+\frac{N_{f}-N_{f_{1}}+N_{f_{2}}}{N_{f}}s&a=1,\ldots,N_{f_{1}}\\
	n_{a}\rightarrow n_{a}-\frac{N_{f}-N_{f_{1}}+N_{f_{2}}}{N_{f}}s&a=1,\ldots,N_{f_{2}}\\
	\tilde\sigma_{i}\rightarrow\tilde\sigma_{i}+\frac{N_{f_{1}}-N_{f_{2}}}{2N_{f}}s&i=1,\ldots,3nN_{f}-N_{c}\\
	\xi\rightarrow\xi-\frac{N_{f_{1}}+N_{f_{2}}}{2}s
\end{cases}
\end{equation}

We study the limit of large $s$ on both the electric (\ref{eq:SUele}) and magnetic (\ref{eq:SUmag}) partition functions by making use of the asymptotic behavior of the hyperbolic Gamma function (\ref{eq:asymptGamma}). We check that the leading saddle point contributions cancel between the electric and magnetic phases, and we are left with the equality between
\begin{equation}
	\mathcal{Z}_{\text{ele}} = \mathcal{Z}_{\text{SU}(N_{c})_{k}}^{(N_{f},N_{a})}(\mu_{a};\nu_{a};\tau_{X};\tau_{Y})
\end{equation}
in which we have no FI since after the flow we integrate on $\xi$, and
\begin{equation}
\begin{split}
	\mathcal{Z}_{\text{mag}} &=  \prod_{j=0}^{n-1}\prod_{\ell=0}^{2}\prod_{a=1}^{N_{f}}\prod_{b=1}^{N_{a}}\Gamma_{h}(j\tau_{X}+\ell\tau_{Y}+\mu_{a}+\nu_{b})\\
	&\times e^{i\pi\phi}\int\dd{\xi}e^{i\pi\eta\xi}\mathcal{Z}_{\text{U}(\tilde{N}_{c})_{-k}}^{(N_{a},N_{f})}(\tau_{X}-\tau_{Y}-\nu_{a}; \tau_{X}-\tau_{Y}-\mu_{a};\tau_{X};\tau_{Y};-\hat\xi)
\end{split}
\end{equation}
where
\begin{equation}
\begin{split}
	\hat\xi &= \xi+\frac{N_{f_{1}}-N_{f_{2}}}{2}(m_{A}-\tau_{X}+\tau_{Y}+\omega),\\
	\eta &=2m_{B}N_{c}+6nm_{A}N_{f}-(1-3nN_{c}+2n^{2})\tau_{X}-6\tau_{Y}+2\omega(1+2n-3nN_{f}).
\end{split}
\end{equation}
From (\ref{eq:SUmagp0}) we can see the level $-1$ mixed CS, in the term $m_{B}\xi$, between the abelian subgroup of the $\text{U}(\tilde{N}_{c})$ and the abelian $\text{U}(1)$ gauge group. In this case, the $\text{U}(1)$ gauge group does not have a CS level. Again the real masses satisfy the constrain $\sum_{a=1}^{N_{f}}\mu_{a}=\sum_{a=1}^{N_{f}}\nu_{a}=N_{f}m_{A}$ where $\mu_{a}=m_{a}+m_{A}$ and $\nu_{a}=n_{a}+m_{A}$.

The complex exponent $\phi$ needed for the matching is given by
\begin{equation}
\begin{split}
	\phi&=9nN_{f}km_{A}^{2}-kN_{c}m_{B}^{2}+\frac{\tau_{X}}{2}(N_{f_{1}}-N_{f_{2}})\Big(N_{c}(1+3n)-3nN_{f}(n+3)\Big)m_{A}\\
	&+N_{c}(N_{f_{2}}-N_{f_{1}})m_{B}\omega+\frac{\tau_{X}^{2}}{2}\Big(nN_{f}(13-9n-18n^{2})+N_{c}(-1-4n+6n^{2})\Big)\\
	&+\frac{3}{2}n(N_{f_{1}}-N_{f_{2}})\sum_{a=1}^{N_{f}}m_{a}^{2}.
\end{split}
\end{equation}

\section{Dualities for $\text{USp}(2N_{c})$ chiral SQCD with two anti-symmetric}
In this section we start by setting up the notation for $\text{USp}(2N_{c})$ theories and their dualities. The partition function of a CS-theory with $\text{USp}(2N_{c})_{k}$ gauge group can be found starting from (\ref{eq:Zgeneral}). In particular, for the case of our interest, with $2N_{f}$ fundamentals and two anti-symmetric rank two tensors $A,B$, the partition function is given by
\begin{equation}
\begin{split}
	\mathcal{Z}_{\text{USp}(2N_{c})_{2k}}^{2N_{f}}(\vec\mu;\tau_{A};\tau_{B})&=\frac{\Gamma_{h}(\tau_{A})^{N_{c}}\Gamma_{h}(\tau_{B})^{N_{c}}}{2^{N_{c}}N_{c}!\sqrt{-\omega_{1}\omega_{2}}^{N_{c}}}\int\prod_{i=1}^{N_{c}}\dd{\sigma_{i}} \exp(-i\pi k\sigma_{i}^{2})\frac{\prod_{a=1}^{2N_{f}}\Gamma_{h}(\mu_{a}\pm \sigma_{i})}{\Gamma_{h}(\pm2\sigma_{i})}\\
	&\times\prod_{1\le i < j \le N_{c}}\prod_{\alpha=A,B}\frac{\Gamma_{h}(\tau_{\alpha}\pm\sigma_{i}\pm\sigma_{j})}{\Gamma_{h}(\pm\sigma_{i}\pm\sigma_{j})}.
\end{split}
\end{equation}
The $3d$ duality for the $\text{USp}(2N_{c})_{0}$ case was worked out in \cite{Amariti:2022iaz} and relates
\begin{itemize}
	\item $3d$ $\mathcal{N}=2$ $\text{USp}(2N_{c})_{0}$ SCQD with $2N_{f}$ flavours $Q$ and two anti-symmetric rank-two tensors $A,B$ interacting through the superpotential
	\begin{equation}
		\mathcal{W}_{\text{ele}} = \Tr A^{n+1}+\Tr AB^{2}.
		\label{eq:Uspele}
	\end{equation}
	\item $3d$ $\mathcal{N}=2$ $\text{USp}(2\tilde{N}_{c})_{0}$ SQCD where $\tilde{N}_{c}=3nN_{f}-N_{c}-2n-1$, with $2N_{f}$ dual flavours $q$, two anti-symmetric rank-two tensors $a,b$ interacting through the superpotential
	\begin{equation}
	\begin{split}
		\mathcal{W}_{\text{mag}}&=\Tr a^{n+1}+\Tr ab^{2}+\sum_{j=0}^{n-1}\sum_{\ell=0}^{2}\mathcal{M}_{j,\ell}qa^{j}b^{\ell}q\\
		&+\sum_{\substack{j=0,\ldots,n-1\\\ell=0,1,2\\j\ell=0}}Y_{j,\ell}\widetilde{Y}_{n-j,2-\ell}+\sum_{q=0}^{\frac{n-3}{2}}Z_{q}\widetilde{Z}_{\frac{n-3}{2}-q}\;.
	\end{split}
	\label{eq:Uspmag}
	\end{equation}
\end{itemize}
The non-anomalous global symmetry of the teories is $\text{SU}(2N_{f})\times\text{U}(1)_{A}\times\text{U}(1)_{R}$ under which the fields transform as in table \ref{tab:USp}.

\begin{table}
\centering
\renewcommand{\arraystretch}{1.2}
\caption{Matter content of $\text{USp}(2N_{c})_{0}$ and $\text{USp}(2\tilde{N}_{c})_{0}$ dual theories. }
\vspace{5pt}
\ytableausetup{centertableaux}
\adjustbox{max width=\columnwidth}{
\begin{tabular}{||c|c|c|c|c|c||}
	\cline{2-6}
	\multicolumn{1}{c|}{}&\multicolumn{2}{c|}{Gauge}&\multicolumn{3}{c||}{Global}\\
	\hline
	Field & USp$(2N_{c})$ & USp$(2\tilde N_{c})$ & SU$(2N_{f})$ & U$(1)_{A}$ & U$(1)_{R}$\\
	\hline
	$Q$ & $\Box$ & $1$ & $\Box$ & $1$ & $r_{Q}$\\
	$A$ & $\ydiagram{1,1}$ & $1$ & $1$ & $0$ & $\frac{2}{n+1}$\\
	$B$ & $\ydiagram{1,1}$ & $1$ & $1$ & $0$ & $\frac{n}{n+1}$\\
	$Y_{j\ell}^{\pm}$ & 1 & $1$ & $1$ & $-2N_{f}$  & $2(1-r_{Q})N_{f}+\frac{2j+n\ell-2(N_{c}+n)}{n+1}$\\
	$Z_{q}^{\pm}$ & 1 & $1$ & $1$ & $-4N_{f}$ & $4(1-r_{Q})N_{f}+\frac{2+4q-4(N_{c}+n)}{n+1}$\\
	\hline\hline
	$q$ & $1$ & $\Box$ & $\overline\Box$ & $-1$ & $\frac{2-n}{n+1}-r_{Q}$\\
	$a$ & $1$ & $\ydiagram{1,1}$ & $1$ & $0$ & $\frac{2}{n+1}$\\
	$b$ & $1$ & $\ydiagram{1,1}$ & $1$ & $0$ & $\frac{n}{n+1}$\\
	$\mathcal{M}_{j,0}^{j=0,\ldots,n-1}$ & $1$ & $1$ & $\ydiagram{1,1}$ & $2$ & $2r_{Q}+\frac{2j}{n+1}$\\
	$\mathcal{M}_{2j,1}^{j=0,\ldots,\frac{n-1}{2}}$ & $1$ & $1$ & $\ydiagram{1,1}$ & $2$ & $2r_{Q}+\frac{4j+n}{n+1}$\\
	$\mathcal{M}_{2j+1,1}^{j=0,\ldots,\frac{n-3}{2}}$ & $1$ & $1$ & $\ydiagram{1,1}$ & $2$ & $2r_{Q}+\frac{4j+n+2}{n+1}$\\
	$\mathcal{M}_{j,2}^{j=0,\ldots,n-1}$ & $1$ & $1$ & $\ydiagram{1,1}$ & $2$ & $2r_{Q}+\frac{2j+2n}{n+1}$\\
	$\tilde Y_{j\ell}^{\pm}$ & 1 & $1$ & $1$ & $2N_{f}$ & $2(r_{Q}-1)N_{f}+\frac{n\ell+2(j+N_{c}+n+1)}{n+1}$\\
	$\tilde Z_{q}^{\pm}$ & $1$ & $1$ & $1$ & $4N_{f}$ & $4(r_{Q}-1)N_{f}+\frac{4(q+N_{c}+n)+6}{n+1}$\\
	\hline
\end{tabular}}
\label{tab:USp}
\end{table}

At the level of the partition function, this duality corresponds to the identity
\begin{equation}
\begin{split}
	\mathcal{Z}_{\text{USp}(2N_{c})}^{2N_{f}}(\mu_{a};\tau_{A};\tau_{B}) &= \mathcal{Z}_{\text{USp}(2\tilde{N}_{c})}^{2N_{f}}(\tau_{A}-\tau_{B}-\mu_{a};\tau_{A};\tau_{B})\\
	&\times\prod_{j=0}^{n-1}\prod_{\ell=0}^{2}\prod_{1\le a<b\le 2N_{f}}\Gamma_{h}(j\tau_{A}+\ell\tau_{B}+\mu_{a}+\mu_{b})\\
	&\times\prod_{q=0}^{\frac{n-3}{2}}\prod_{a=1}^{2N_{f}}\Gamma_{h}((2q+1)\tau_{A}+\tau_{B}+2\mu_{a})\\
	&\times\prod_{\substack{j=0,\ldots,n-1\\\ell=0,1,2\\j\ell=0}}\Gamma_{h}\qty(j\tau_{A}+\ell\tau_{B}+2N_{f}\omega-(N_{c}+n)\tau_{A}-\sum_{a=1}^{2N_{f}}\mu_{a})\\
	&\times\prod_{q=0}^{\frac{n-3}{2}}\Gamma_{h}\qty((2q+1)\tau_{A}+4N_{f}\omega-2(N_{c}+n)\tau_{A}-2\sum_{a=1}^{2N_{f}}\mu_{a})
\end{split}
\label{eq:Uspdual}
\end{equation}
The superpotential (\ref{eq:Uspele}) fixes the values of the real masses for the adjoint fields 
\begin{equation}
	\tau_{A}=\frac{2\omega}{n+1},\qquad \tau_{B}=\frac{n \omega}{n+1}.
\end{equation}
This duality is going to be our starting point.

In the following we want to construct the duality for non-vanishing CS level. We consider the duality without CS terms and
with $2(N_{f}+k)$ fundamental flavours and assign a positive real mass to $2k$. By integrating out the massive fields
we arrive at the duality between
\begin{itemize}
	\item $\text{USp}(2N_{c})_{2k}$ SQCD with $2N_{f}$ fundamentals $Q$ and two anti-symmetric rank-two tensors $A,B$ interacting through the superpotential
	\begin{equation}
		\mathcal{W}_{\text{ele}}=\Tr A^{n+1}+\Tr AB^{2}.
	\end{equation}
	\item $\text{USp}(2\tilde{N_{c}})$ SQCD with $\tilde{N}_{c}=3n(N_{f}+\abs{k})-N_{c}-2n-1$,  $2N_{f}$ fundamentals $q$ and two anti-symmetric rank-two tensors $a,b$ interacting through the superpotential
	\begin{equation}
		\mathcal{W}_{\text{mag}} = \Tr a^{n+1}+\Tr ab^{2}+\sum_{j=0}^{n-1}\sum_{\ell=0}^{2}\mathcal{M}_{j,\ell}qa^{j}b^{\ell}q.
	\end{equation}
	The electric and the magnetic theories acquire a CS level $2k$ and $-2k$ respectively
\end{itemize}
The dressed monopole operators of the electric theory acting as singlets  in the dual phase become massive and are integrated out.

To reproduce this flow at the level of the partition function, we start from the equality (\ref{eq:Uspdual}) and consider the following shifts in the real masses
\begin{equation}
\begin{cases}
	m_{A}\rightarrow m_{A}+\frac{k}{N_{f}+k}s\\
	m_{a}\rightarrow m_{a}-\frac{k}{N_{f}+k}s&a=1,\ldots,2N_{f}\\
	m_{a}\rightarrow m_{a}+\frac{N_{f}}{N_{f}+k}s&a=1,\ldots,2k\\
\end{cases}
\end{equation}

We study the limit of large $s$ on both sides of the identity (\ref{eq:Uspdual}) by making use of the asymptotic behavior of the hyperbolic Gamma function (\ref{eq:asymptGamma}). We check that the leading saddle point contributions cancel between the electric and magnetic phases, and we are left with the equality between
\begin{equation}
	\mathcal{Z}_{\text{ele}}=\mathcal{Z}_{\text{USp}(2N_{c})_{2k}}^{2N_{f}}(\mu_{a};\tau_{A};\tau_{B}),
\end{equation}
and
\begin{equation}
\begin{split}
	\mathcal{Z}_{\text{mag}}&=e^{i\pi\phi}\mathcal{Z}_{\text{USp}(2\tilde{N}_{c})_{-2k}}^{2N_{f}}(\tau_{A}-\tau_{B}-\mu_{a};\tau_{A};\tau_{B})\\
	&\times\prod_{j=0}^{n-1}\prod_{\ell=0}^{2}\prod_{1\leq a < b \leq 2N_{f}}\Gamma_{h}(j\tau_{A}+\ell\tau_{B}+\mu_{a}+\mu_{b}).
\end{split}
\end{equation}
To evaluate the asymptotic behavior of the factor coming from the singlets, or the electric mesons under the duality map, in the magnetic theory, we make use of the following decomposition formula
\begin{equation}
\begin{split}
	\sum_{\substack{a,b=1,\ldots,2k\\a<b}}(j\tau_{A}+\ell\tau_{B}+\mu_{a}+\mu_{b}-\omega) &= 2(k-1)\sum_{a=1}^{2k}\mu_{a}^{2}+\qty(\sum_{a=1}^{2k}\mu_{a})^{2}\\
	&+(4k-2)(j\tau_{A}+\ell\tau_{B}-\omega)\sum_{a=1}^{2k}\mu_{a}\\
	&+k(2k-1)(j\tau_{A}+\ell\tau_{B}-\omega)^{2}.
\end{split}
\end{equation}

The complex exponent $\phi$ necessary for the equality between the partition functions to hold has the following form
\begin{equation}
\begin{split}
	\phi&=12nN_{f}(k-N_{f})m_{A}^{2}+6\tau_{A}\Big(-1-2n-2N_{c}+2N_{f}(1+n)+2k(-1+2n)\Big)m_{A}\\
	&+\frac{\tau_{A}^{2}}{24}(6k-3)+\frac{\tau_{A}\tau_{B}}{12}\Big(4k-12k^{2}(1+24n^{2})+3\big(-7-25n(1+n)-24N_{f}^{2}(1+n)^{2}\\
	&-24N_{c}(1+2n)-24N_{c}^{2}+24N_{f}(1+n)(1+2n+2N_{c})\\
	&+4k(12N_{f}-72nN_{f}(1+n)+n(39+77n+54N_{c}))\big)\Big)\\
	&-3n\qty(\qty(\sum_{a=1}^{2N_{f}}m_{a})^{2}-2k\sum_{a=1}^{2N_{f}}m_{a}^{2}).
\end{split}
\end{equation}

\section{Conclusions}

In this paper we studied $3d$ $\mathcal{N}=2$ dualities for two adjoint $\text{U}(N_{c})$ and $\text{SU}(N_{c})$ SQCD with $D_{n+2}$-type superpotential and odd $n$.
This generalizes the  constructions of \cite{Benini:2011mf,Aharony:2014uya} for  SQCD and of  \cite{Hwang:2015wna,Nii:2019qdx,Amariti:2020xqm} for adjoint SQCD.
The dualities are obtained starting from the one obtained in \cite{Hwang:2018uyj} from the $4d/3d$ reduction.
The classification is constructed through real mass flows, Higgs flows and the gauging of the topological symmetry.
We corroborated these construction by checking the various steps with the help of the three sphere partition function.
Furthermore we matched the CS contact terms across the dual phases with the complex phases that can be read in the integral identities on the three sphere.
We concluded by proposing a duality for $\text{USp}(2N)_{2k}$ SQCD with tow antisymmetric and $D_{n+2}$ type superpotential 
that was overlooked in the literature.

There are interesting aspects of such dualities and possible generalizations that we did not investigate and that we leave for future projects.
For example we did not match the superconformal index across the new dual phases. This should provide a stronger check of the dualities obtained here. 
Another aspect that we did not investigate corresponds to find mirror dualities for the $\text{U}(1)$ sectors in the duals of the $\text{SU}(N_c)$ dualities in presence of charged matter fields. Such mirrors could simplify the structure of the dual models, that so far are given in terms of product groups.
A further generalization of the construction is related to chiral models with monopole superpotentials. In the SQCD case such possibility has been discussed in \cite{Benini:2017dud} for the case of linear monopole superpotential and in 
\cite{Amariti:2018gdc} for SQCD with quadratic monopole superpotential. In the $A_{n}$ case a similar extension (with quadratic monopoles) has been proposed in \cite{Amariti:2019rhc}.
We conclude observing that a full list of $4d$ dualities for $\text{SU}(N_{c})$ SQCD with two tensors and monopole superpotentials
has been provided in \cite{Brodie:1996xm}. It should be possible to reduce such dualities to 3d and that to study the chiral limiti of these cases as well. This is an interesting possibilities because some of the models in the classification have
also an interpretation in terms of the HW setup \cite{Brunner:1998jr}. This may allow to study the $4d/3d$ reduction
from a T-duality in the HW setup along the lines of \cite{Amariti:2015mva,Amariti:2016kat} and the chiral dualities as discussed in \cite{Amariti:2020xqm}

%
%
%
%
%
%
\section*{Acknowledgments}
%
%
This work has been supported in part by the Italian Ministero dell'Istruzione, 
Universit\`a e Ricerca (MIUR), in part by Istituto Nazionale di Fisica Nucleare (INFN) through the “Gauge Theories, Strings, Supergravity” (GSS) research project and in part by MIUR-PRIN contract 2017CC72MK-003.  

\bibliographystyle{JHEP}
\bibliography{Brodie.bib}

\end{document}